\documentclass[lettersize,journal]{IEEEtran}

\PassOptionsToPackage{hyphens}{url}

\usepackage{setspace}
\usepackage{tabularx}

\usepackage{amsmath}
\usepackage{amsthm}
\usepackage{amssymb}    

\usepackage{makecell}
\usepackage{longtable}
\usepackage{booktabs}
\usepackage{float}
\usepackage[table]{xcolor}   
\usepackage{multirow}
\usepackage{stfloats}

\usepackage[ruled,vlined]{algorithm2e}

\usepackage{graphicx}
\usepackage{subfig}
\usepackage{url}
\usepackage{cite}
\usepackage{textcomp}

\usepackage[skins,breakable]{tcolorbox}
\usepackage{enumitem}

\newtcolorbox{lessonslearned}[1][Lessons Learned]{
    enhanced,
    breakable,
    width=\linewidth,
    colback=white,
    colframe=black,
    boxrule=0.85pt,
    arc=0pt,
    outer arc=0pt,
    left=7pt,
    right=7pt,
    top=10pt,
    bottom=6pt,
    before skip=19pt,
    after skip=8pt,
    fontupper=\small,
    overlay unbroken and first={
        \node[
            anchor=south west,
            fill=black,
            text=white,
            draw=black,
            line width=0.85pt,
            inner xsep=9pt,
            inner ysep=3.2pt,
            font=\bfseries\footnotesize
        ] at ([xshift=-0.85pt,yshift=-0.85pt]frame.north west)
        {#1};
    }
}

\newlist{lessonitems}{itemize}{1}
\setlist[lessonitems]{
    label=\textbullet,
    leftmargin=1.55em,
    labelsep=0.55em,
    topsep=5pt,
    itemsep=4pt,
    parsep=0pt,
    partopsep=0pt
}

\usepackage{bbding}
\usepackage{verbatim}
\usepackage{flushend}

\usepackage[
    colorlinks,
    linkcolor=red,
    anchorcolor=green,
    citecolor=blue,
    urlcolor=blue
]{hyperref}

\usepackage{cleveref}

\crefname{figure}{Fig.}{Figs.}
\Crefname{figure}{Fig.}{Figs.}
\crefname{table}{Table}{Tables}
\Crefname{table}{Table}{Tables}
\crefname{section}{Section}{Sections}
\Crefname{section}{Section}{Sections}

\begin{document}

\title{High-Altitude Platforms Beyond Connectivity: A Survey of Integrated Sensing, Storage, Communication, Computing, and Intelligence}

\author{Haoxiang Luo, Mohamed-Slim Alouini,~\IEEEmembership{Fellow,~IEEE}
\thanks{H. Luo and M. S. Alouini are with the Computer, Electrical and Mathematical Science and Engineering (CEMSE) Division in King Abdullah University of Science and Technology (KAUST), Thuwal 6900, Makkah Province, Saudi Arabia. (emails: \{haoxiang.luo, slim.alouini\}@kaust.edu.sa).
}

}



\maketitle

\begin{abstract}

High-altitude platforms (HAPs) are emerging as persistent middle-layer infrastructures for space-air-ground integrated networks (SAGINs), offering a favorable compromise among coverage, latency, endurance, and deployment flexibility. Their role, however, is evolving beyond communication relaying toward the joint provision of sensing, storage, communication, computing, and intelligence (S$^2$C$^2$I). 
This survey presents a unified HAP-centric perspective on S$^2$C$^2$I integration. We first review HAP fundamentals, platform categories, and their principal roles in SAGINs, including wide-area access, relaying, backhaul, edge service, low-altitude aerial coordination, and cross-layer orchestration. We then develop an integrated architecture spanning multi-plane connectivity, payload functional splits, and a cloud-edge-HAP space continuum with hierarchical data, control, computing, and storage loops. The enabling technologies are systematically examined, covering heterogeneous RF, millimeter-wave, terahertz, free-space optical, and hybrid links; sensing payloads and integrated sensing and communication; onboard computing; storage and caching; and AI-based orchestration. We further synthesize standardization progress, open software and datasets, testbeds, field evidence, and a four-level evaluation methodology ranging from component validation to mission-level effectiveness. An emergency-response case study demonstrates that joint S$^2$C$^2$I orchestration substantially improves conjunctive service availability while reducing feeder-link traffic. 
Finally, we identify research opportunities in agentic AI, trustworthy autonomy, goal-oriented semantic operation and digital twins, and sustainable, certifiable, and open HAP-native systems. The resulting synthesis provides a coherent roadmap from platform design to network-wide deployment.

\end{abstract}


\section{Introduction} \label{sec-I}

\subsection{Background and Motivation}

\IEEEPARstart{N}{on}-Terrestrial Networking (NTN) has evolved from a satellite-centric extension of terrestrial coverage into a standardized component of 5G-Advanced and 6G \cite{wang2025toward}. In the 3GPP view \cite{3gpp38811}, NTN includes both satellites and airborne platforms, with airborne vehicles typically operating between 8 and 50 km. In comparison, the ITU radio regulations define High-Altitude Platforms (HAPs) more specifically as radio stations located on objects at altitudes of 20–50 km \cite{kurt2021vision}. Importantly, 3GPP also distinguishes transparent and regenerative payload modes, meaning that an airborne platform can range from a bent-pipe relay to a node with onboard demodulation, decoding, routing, and base-station functions. These definitions are not merely terminological; they determine how much sensing, computation, and storage can be embedded directly on a HAP. 

Among NTN platforms, HAPs occupy a particularly attractive middle layer. Compared with satellites, HAPs are closer to the ground and enable lower-latency, more flexible deployment \cite{alam2021high}; compared with low-altitude Unmanned Aerial Vehicles
(UAVs), they offer wider-area coverage, longer endurance, and a more stable service footprint \cite{almalki2022incorporating, wang2026optimization}. Recent HAPS-oriented surveys therefore position HAPs as bridges between terrestrial and non-terrestrial infrastructure \cite{lou2023haps}, \cite{jamshed2026integrated}, supporting last-mile access, integrated access and backhaul, aerial relaying, and cross-layer control functions in future 6G systems. 

At the same time, the functional role of HAPs is expanding far beyond communication coverage. Representative primary works have proposed HAP-enabled edge computing for massive Internet of Things (IoT) connectivity \cite{alotaibi2025optimizing}, communication-computing-sensing architectures for aerial delivery systems \cite{gao2024cost}, and HAPs-assisted caching and computation offloading for intelligent transportation systems \cite{ren2022caching}. More recently, HAPs have also been examined as platforms where free-space optical communications and atmospheric optical sensing can share the same optical and power envelope \cite{trinh2025optical}. Due to these characteristics, HAPs are regarded as an important means of communication that can cover unconnected areas \cite{lin2026haps}, including deserts, plateaus, and oceans, etc. Taken together, these developments indicate that HAPs are increasingly being treated as persistent airborne service infrastructures capable of jointly Sensing, Storage, Communication, Computing,  and Intelligence (S$^2$C$^2$I) \cite{le2026simultaneous}, \cite{lin2025connectivity}, as shown in Fig. \ref{fig:func}. Specifically, the reasons why HAPs need to integrate these functions are as follows:

 \begin{figure}[!t]
\centering
 \includegraphics[width=3.5 in]{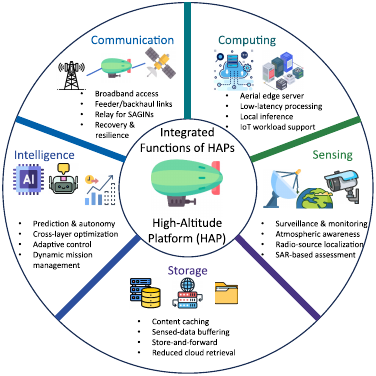}
   \caption{Integrated sensing, storage, communication, computing, and intelligence functions of HAPs.}
\label{fig:func}
   \vspace{-0.5cm}
\end{figure}

\begin{itemize}
    \item The need for \textbf{sensing} on HAPs is both technical and application-driven. On the application side, HAPs are increasingly considered for maritime or border surveillance, atmospheric monitoring, radio-source localization, and Synthetic Aperture Radar (SAR)-based disaster assessment \cite{yusdian2023concept}, \cite{huang2025design}. On the system side, sensing also underpins safe and efficient HAPs operation itself \cite{benaya2025aerial}. Platform state estimation, environmental awareness, interference perception, beam adaptation, and mission-context awareness all depend on timely observation of the physical and radio environment. 
    \item The need for \textbf{storage} is equally important and is often understated in current surveys. Storage on HAPs supports proactive and reactive content caching, buffering of sensed data, and reduction of repeated cloud retrieval over constrained feeder links \cite{ren2022caching}, \cite{vallero2022caching}. 3GPP NTN service discussions already acknowledge the importance of content delivery optimization and, in newer phases, store-and-forward support for delay-tolerant operation \cite{3gpp38811}, \cite{qiu2022mobile}. 
    \item The need for \textbf{communication} is obvious but should be interpreted more broadly than access connectivity alone. 3GPP NTN  use cases and the ITU HAPs agenda encompass broadband access, feeder links, relay/backhaul support, network resilience, multicast/broadcast, and disaster recovery \cite{3gpp38811}. In HAPs-specific studies, the platform is also treated as a network extension layer for SAGIN \cite{niu2024hap}. It can be a coordinator for lower-altitude UAV tiers and a programmable node in future multi-layer topologies. 
    \item The need for \textbf{computing} on HAPs arises because forwarding all raw data to remote clouds or terrestrial cores is often too costly in latency and bandwidth. Recent works explicitly position HAPs as aerial edge servers \cite{yu2024joint}, or collaborative computing nodes for massive IoT and hierarchical aerial computing \cite{ke2021edge}. In such scenarios, onboard HAPs computation reduces round-trip delay. This supports feature extraction and local inference, mitigates handoff-induced service disruption \cite{liu2024distributed}. 
    \item Finally, the need for \textbf{intelligence} follows from the non-stationary, cross-layer nature of HAPs operation. HAPs must operate under time-varying traffic demands, energy budgets, atmospheric uncertainty, and heterogeneous service requirements \cite{bakambekova2024interplay}, \cite{shen2023survey}. Static rule-based control is increasingly insufficient for such environments. Artificial Intelligence (AI) enables prediction, autonomy, and joint optimization across other functions \cite{luo2026ai}.  Accordingly, recent studies investigate AI methods for HAPs \cite{xing2026generative}, \cite{dahrouj2023machine}. 
\end{itemize}

As a result, S$^2$C$^2$I integration on HAPs is not a simple aggregation of five independent capabilities. Instead, it is a tightly coupled co-design problem. Sensing payloads produce data streams that affect communication and storage loads; storage trades off with backhaul usage and cache hit rate; computing changes the timing and volume of data transport; and intelligence must arbitrate among energy, payload, and QoS objectives under uncertain conditions \cite{ke2021edge}, \cite{ren2022caching}, \cite{vallero2022caching}, \cite{qiu2022mobile}. This coupling is even more pronounced in stratospheric platforms because power is limited, endurance is mission-critical, maintenance cycles are long, and payload mass/volume are expensive resources \cite{dhruv2025review}. Consequently, a survey that studies these functions separately risks missing the true system bottlenecks and design opportunities.

\begin{table*}[!t]
\centering %
    \centering
    \caption{Comparison of HAPs Surveys From an S$^2$C$^2$I Perspective}
    \label{tab:related_works}
    
    \renewcommand{\arrayrulewidth}{0.8pt} 
    \renewcommand{\tabcolsep}{10pt} 
    
    {\fontsize{8}{10}\selectfont 
     
    \begin{tabular}{m{0.5cm}<{\centering}||m{0.5cm}<{\centering}|m{1cm}<{\centering}|m{1cm}<{\centering}|m{1cm}<{\centering}|m{1cm}<{\centering}|m{1cm}<{\centering}|m{1.5cm}<{\centering}|m{3.9cm}<{\centering}} 
        \hline
        \hline
        \rowcolor{gray!15}
         \textbf{Ref.} & \textbf{Year} &  \textbf{Sensing} & \textbf{Storage} & \textbf{Comm.} & \textbf{Comp.} & \textbf{Intell.} & \textbf{Cross-Function} & \textbf{Primary Scope}\\ 
         \hline
        \rowcolor{blue!10}
         \cite{karapantazis2005broadband} & 2005 & \XSolidBrush & \XSolidBrush & \Checkmark & \XSolidBrush & \XSolidBrush & \XSolidBrush & Broadband communication, spectrum, and access techniques\\
        \hline
        \rowcolor{blue!10}
        \cite{widiawan2007high} & 2007 & \XSolidBrush & \XSolidBrush & \Checkmark & \XSolidBrush & \XSolidBrush & \XSolidBrush & HAPs communication infrastructure and applications\\
        \hline
        \rowcolor{blue!10}
         \cite{arum2020review} & 2020 & \XSolidBrush & \XSolidBrush & \Checkmark & \XSolidBrush & $\bigcirc$ & \XSolidBrush & Rural coverage extension and terrestrial coexistence\\
        \hline
        \rowcolor{blue!10}
         \cite{kurt2021vision} & 2021 & $\bigcirc$ & $\bigcirc$ & \Checkmark & \Checkmark & \Checkmark & $\bigcirc$ & Future HAPs networks, payloads, caching, and AI\\
        \hline
        \rowcolor{green!10}
         \cite{lou2023haps} & 2023 & \XSolidBrush & \XSolidBrush & \Checkmark & \XSolidBrush & \XSolidBrush & \XSolidBrush & Prospective HAPs-enabled NTN architectures\\
        \hline
        \rowcolor{green!10}
         \cite{abbasi2024haps} & 2024 & \XSolidBrush & \XSolidBrush & \Checkmark & $\bigcirc$ & \XSolidBrush & \XSolidBrush & HAPs use cases, open challenges, and integration with 6G\\
        \hline
        \rowcolor{green!10}
         \cite{svistunov2025bridging} & 2025 & $\bigcirc$ & \XSolidBrush & \Checkmark & $\bigcirc$ & $\bigcirc$ & \XSolidBrush & HAPs use cases, enabling technologies, and NTN\\
        \hline
        \rowcolor{green!10}
         \cite{cao2026survey} & 2026 & \Checkmark & \XSolidBrush & \Checkmark & $\bigcirc$ & $\bigcirc$ & $\bigcirc$ & Near-space channel modeling, transmission protocols\\
        \hline
        \rowcolor{purple!10}
         \cite{bakambekova2024interplay} & 2024 & \XSolidBrush & $\bigcirc$ & \Checkmark & $\bigcirc$ & \Checkmark & $\bigcirc$ & AI-enabled SAGIN optimization, including HAPs\\
        \hline
        \rowcolor{purple!10}
         \cite{elkhazraji2025haps} & 2025 & \Checkmark & \XSolidBrush & \Checkmark & \XSolidBrush & \XSolidBrush & $\bigcirc$ & Joint optical communication and atmospheric sensing\\
        \hline
        \rowcolor{purple!10}
         \cite{ccougay2026hap} & 2026 & \Checkmark & $\bigcirc$ & \Checkmark & \Checkmark & $\bigcirc$ & $\bigcirc$ & HAPs applications in aerial communication\\
        \hline
        \rowcolor{purple!10}
         \cite{huang2026high} & 2026 & $\bigcirc$ & $\bigcirc$ & \Checkmark & \Checkmark & \Checkmark & $\bigcirc$ & HAPS-enabled low-altitude economy and regulation\\
        \hline
        \rowcolor{teal!20}
        \textbf{Ours} & \textbf{2026} & \textbf{\Checkmark} & \textbf{\Checkmark} & \textbf{\Checkmark} & \textbf{\Checkmark} & \textbf{\Checkmark} & \textbf{\Checkmark} & Unified HAPs-centric S$^2$C$^2$I architecture and cross-layer co-design\\
        \hline
        \hline
    \end{tabular}}
 \vspace{2pt} 
\footnotesize{$\bigcirc$: Partially mentioned; \Checkmark: Fully supported; \XSolidBrush: Not supported.}

    \label{tab:related_works}
\end{table*}

\subsection{Related Work and Our Contributions}
The literature related to HAPs has evolved from communication-oriented studies toward visions involving integrated NTNs, sensing, edge computing, and AI. Nevertheless, existing surveys differ considerably in their system boundaries and in the functional roles assigned to HAPS. For clarity, we classify the relevant surveys into three categories, as shown in Table~\ref{tab:related_works}. 

\subsubsection{Communication-Centric Foundations and HAPs Network Design} 

The first category covers foundational studies framing HAPs as airborne communication infrastructure. Early surveys by Karapantazis and Pavlidou \cite{karapantazis2005broadband} and Widiawan and Tafazolli \cite{widiawan2007high} established the basic HAPs communication model by reviewing platform types, frequency allocations, propagation characteristics, antennas, and representative deployments. They positioned HAPs as a compromise between terrestrial Base Stations (BSs) and satellite systems. Bound by the communication demands and technological maturity of the era, these works gave limited attention to onboard data processing and autonomous service provisioning.
Arum et al. \cite{arum2020review} later examined HAPs wireless coverage for rural and underserved areas. It focused on HAPs-terrestrial coexistence, coverage extension, antenna design, interference mitigation, and intelligent radio resource management. Karabulut et al. \cite{kurt2021vision} presented a broader vision of future HAPs networks. It spanned energy and payload systems, radio resource management, waveforms, handover, and the super-macro base-station paradigm.
Despite this progress, communication remains the core organizing dimension of these surveys. 

\subsubsection{HAPs in Integrated NTN and Near-Space Architectures}
The second category studies HAPs through their architectural role in integrated terrestrial and non-terrestrial networks. Lou et al. \cite{lou2023haps} investigated HAPs roles in prospective NTN architectures. They included ad hoc aerial networking, cell-free operation, and integrated access and backhaul. Abbasi et al. \cite{abbasi2024haps} discussed representative HAPs use cases in 6G vertical heterogeneous networks and summarized associated integration, interference, backhaul, and deployment challenges.
More recently, Svistunov et al. \cite{svistunov2025bridging} provided a comprehensive overview of HAPs use cases, enabling technologies, terrestrial-NTN integration, energy models, field trials, and research challenges. Cao et al. \cite{cao2026survey} focused on the broader near-space information network setting, systematically reviewing channel modeling, physical-layer transmission, medium-access protocols, routing, and network management for HAPs and UAVs.
These works clarify HAPs positioning in future multi-layer networks and their integration with terrestrial, satellite, and low-altitude aerial segments. However, their primary abstraction remains that of a communication, relay, access, or networking node. 

\subsubsection{Function-Oriented Convergence: Sensing, Computing, and Intelligence}
The third category moves beyond connectivity to explore integrated HAPs-enabled functions. Bakambekova et al. \cite{bakambekova2024interplay} surveyed the interplay between AI and SAGINs, including AI-assisted HAPs placement, resource allocation, and computation offloading. Nevertheless, HAPs constitute only one component of the broader SAGIN architecture, and the survey is organized around AI algorithms and network optimization rather than the HAPs.
Elkhazraji et al. \cite{elkhazraji2025haps} presented a tutorial on integrating free-space optical communication and atmospheric optical sensing on HAPs, highlighting shared optical hardware, wavelength resources, and power constraints. This work demonstrates function-level integration, but its scope is specialized to optical communication and atmospheric sensing. Çoğay et al. \cite{ccougay2026hap} categorized HAPs applications into advanced aerial communication, integrated sensing, and aerial computing. Huang et al. \cite{huang2026high} further positioned HAPs as communication, computing, sensing, caching, and regulatory hubs for the low-altitude economy.

These recent studies reveal an evident evolution of HAPs from simple airborne relays to versatile multi-functional service infrastructures, yet existing research still suffers from prominent limitations. Current functional designs are highly application-specific and tailored for individual scenarios including optical atmospheric sensing, SAGIN optimization, near-space networking and low-altitude aviation. Specifically, storage is merely embedded as a subsidiary part of caching or edge computing rather than an independent core capability, while onboard intelligence is narrowly regarded as a pure optimization tool instead of a comprehensive module supporting model inference, machine learning and closed-loop autonomous control. Furthermore, existing surveys are fragmented and siloed, focusing merely on communication modules, network architectures or single standalone functions. Simply collocating these separated functional modules cannot achieve genuine S$^2$C$^2$I integration. To fill this critical research gap, we establish a holistic S$^2$C$^2$I system architecture and unify scattered standalone functional technologies under integrated cross-layer HAP network design.
The main contributions of this survey are summarized as follows: 

\begin{itemize} 
\item \textbf{A systematic characterization of HAPs in SAGINs:} This paper firstly analyzes the fundamental features of HAPs in SAGINs. We provide an overview of typical HAP platforms and compare HAPs with satellites, low-altitude UAVs, and terrestrial communication infrastructure. Furthermore, we summarize core roles of HAPs in multi-tier SAGINs: access, relays, backhaul, edge service, aerial coordination, and intermediate orchestration.

\item \textbf{A unified architecture for S$^2$C$^2$I-integrated HAPs:} We propose a HAPs-oriented architecture that integrates S$^2$C$^2$I as mutually dependent capabilities. This architecture illustrates detailed data flows, control loops, and cross-network interactions among space, air, and ground. We also explore the internal coupling relationship of all S$^2$C$^2$I functions under limited onboard payload, energy, spectrum, computing and endurance resources.

\item \textbf{A review of enabling technologies for HAPs:} We classify key enabling technologies for S$^2$C$^2$I-integrated HAPs. 
We mainly focus on cross-functional coordination technologies such as joint resource allocation, space-air-ground collaborative computing, and network orchestration. In addition, we clarify the maturity of individual functions and technical dependencies affecting overall system performance.

\item \textbf{A synthesis of standardization, evaluation, and future research:} We summarize existing standardization progress, experimental platforms, and evaluation methods for HAPs in SAGINs. We further establish a performance metric system and validate joint cross-functional evaluation through a case study showing the advantages of S$^2$C$^2$I-integrated designs. Finally, we identify current limitations and outline future directions in intelligent orchestration, security, interoperability, and practical deployment.
\end{itemize}

\subsection{Organization of This Survey}
\begin{figure*}[!t]
\centering
 \includegraphics[width=5.5 in]{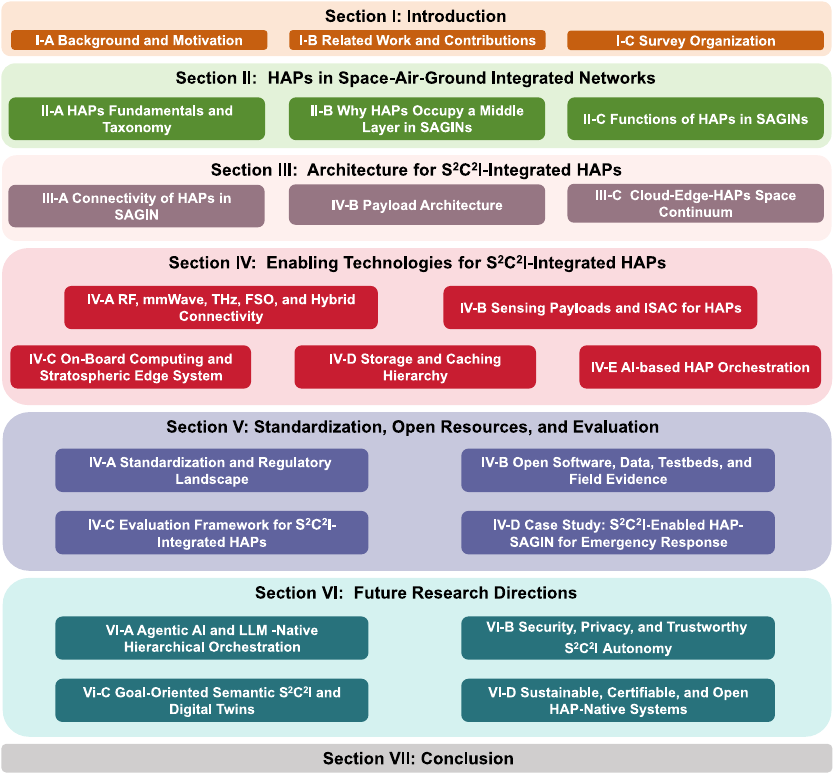}
   \caption{Organization of our survey.}
\label{fig:out}
   \vspace{-0.5cm}
\end{figure*}

The remainder of this survey is organized as shown in Fig. \ref{fig:out} and as follows. Section~\ref{Sec-II} introduces HAPs in SAGINs. Section~\ref{Sec-III} presents the architecture for S$^2$C$^2$I-integrated HAPs. Section~\ref{Sec-IV} reviews the corresponding enabling technologies. Section~\ref{Sec-V} summarizes standardization, open resources, and evaluation methods. Section~\ref{Sec-VI} discusses future research directions, and Section~\ref{Sec-VII} concludes the survey.

\section{HAPs in Space-Air-Ground Integrated Networks}\label{Sec-II} 

\subsection{HAPs Fundamentals and Taxonomy}

HAPs are formally defined by the ITU as radio stations mounted on airborne objects operating at altitudes between 20 km and 50 km in the stratosphere, typically maintaining a quasi-stationary position relative to the Earth's surface \cite{itu2025haps}. The 3GPP further expands this concept in its NTN specifications, classifying HAPs as airborne infrastructure supporting both transparent bent-pipe relaying and regenerative baseband processing payloads \cite{3gpp38811}. Operating above commercial air traffic and most meteorological disturbances, HAPs combine wide coverage with low latency and deployment flexibility, making them a core building block of next-generation information infrastructure.
As shown in Table \ref{tab:hap_taxonomy}, from an aerodynamic perspective, HAPs can be broadly categorized into two major families: Lighter-Than-Air (LTA) and Heavier-Than-Air (HTA) systems, each with distinct technical characteristics and application scenarios \cite{kurt2021vision, svistunov2025bridging}.

\subsubsection{LTA Platforms}
LTA platforms generate lift via buoyant gas \cite{elshazly2025design}, typically helium, and require minimal propulsion to maintain station-keeping. The two primary subtypes are stratospheric airships and high-altitude balloons.
\begin{itemize}
    \item \textbf{Stratospheric airships} feature rigid aerodynamic envelopes, propulsion systems, and attitude control mechanisms, enabling long-duration station-keeping at fixed geographic positions. Representative models include Thales Alenia Space's Stratobus, designed for 12-month endurance at 20 km altitude with a 250 kg payload capacity \cite{thales2025stratobus}, and China's Yuanmeng airship, which completed its maiden stratospheric flight in 2015 \cite{china2015yuanmeng}. Airships excel in persistent surveillance, regional communication coverage, and atmospheric monitoring missions requiring long-term fixed-point presence.
    \item \textbf{High-altitude balloons} are unpowered or minimally powered buoyant systems that drift with stratospheric wind layers \cite{wang2025influence}. The most prominent example is Google's Loon project, which utilized steerable helium balloons at 18-25 km altitude to deliver communication coverage over remote areas \cite{loon2021final}. While lacking precise station-keeping capability, balloon systems offer extremely low deployment costs and are well-suited for disaster response scenarios.
\end{itemize}

\subsubsection{HTA Platforms}
HTA platforms generate lift via aerodynamic forces from fixed wings, relying on solar or hydrogen power for extended endurance \cite{riccio2026preliminary}.
\begin{itemize}
    \item \textbf{Solar-powered fixed-wing UAVs} are the most widely studied HTA HAP platform. Airbus developed the Zephyr series \cite{aalto2025zephyr}, which holds the world record for longest unmanned flight at 64 days of continuous stratospheric operation, with a service ceiling of 21 km and a payload capacity of approximately 5 kg. SoftBank's Hawk30 platform \cite{softbank2024hawk30}, developed in partnership with AeroVironment, targets 24/7 regional connectivity with a 20 km cruising altitude and larger payload envelope.
    \item \textbf{Hydrogen-powered UAVs}, such as Boeing's Phantom Eye \cite{boeing2022phantom}, use liquid hydrogen fuel to achieve multi-day endurance at altitudes up to 20 km. It offers higher payload capacity than solar-powered alternatives at the cost of periodic refueling requirements.
\end{itemize}

\begin{table}[!t]
\centering
\caption{Taxonomy and Key Specifications of Typical HAPs}
\label{tab:hap_taxonomy}

\renewcommand{\arrayrulewidth}{0.8pt}
\renewcommand{\tabcolsep}{5.5pt}

{\fontsize{7.5}{9.5}\selectfont 
\begin{tabular}{m{1.6cm}<{\centering}|m{1.4cm}<{\centering}|m{1.2cm}<{\centering}|m{1.3cm}<{\centering}|m{1.2cm}<{\centering}}
\hline\hline
\rowcolor{gray!15}
\textbf{Category} & \textbf{Model} & \textbf{Altitude} & \textbf{Endurance} & \textbf{Payload}\\
\hline
& Stratobus & 20 km & 12 months & 250 kg\\
\cline{2-5}
\multirow{-2}{*}{LTA Airship} & Yuanmeng & 20–24 km & Days–months & 100 kg\\
\hline
LTA Balloon & Loon System & 18–25 km & 100+ days & 10 kg\\
\hline
& Zephyr S & 18–21 km & 64+ days & 5 kg\\
\cline{2-5}
\multirow{-2}{*}{HTA Solar UAV} & Hawk30 & 20 km & Months & 15 kg\\
\hline
HTA Hydrogen & Phantom Eye & 20 km & 4–10 days & 200 kg\\
\hline\hline
\end{tabular}}
\vspace{2pt}
\end{table}

 \begin{figure*}[!t]
\centering
 \includegraphics[width=6 in]{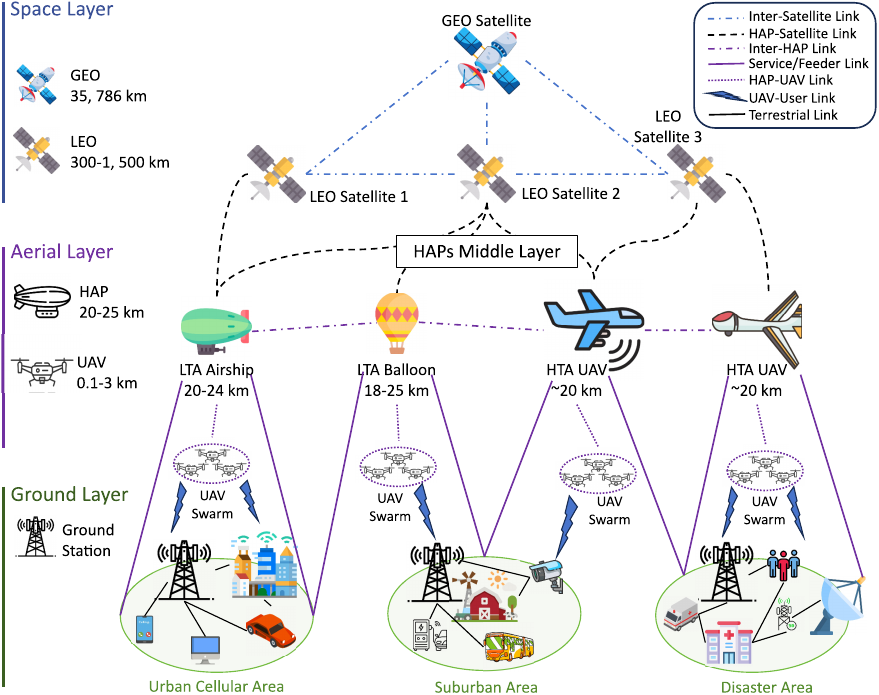}
   \caption{HAPs in the SAGIN architecture. The three-tier SAGIN architecture, where HAPs serve as the pivotal middle layer bridging the space layer, aerial layer, and ground layer with diverse inter-layer communication links.}
\label{fig:topo}
   \vspace{-0.5cm}
\end{figure*}

\subsection{Why HAPs Occupy a Middle Layer in SAGINs}

SAGINs adopt a three-tier hierarchical architecture consisting of the space layer, including Geostationary Earth Orbit (GEO) and Low-Earth-Orbit (LEO) satellites, the aerial layer such as HAPs and UAVs, and the ground layer \cite{yu2026topology}, \cite{hu2025generative}, as shown in Fig. \ref{fig:topo}. Within this architecture, HAPs occupy a unique intermediate stratum that bridges the gap between orbital satellites and terrestrial/aerial infrastructure, addressing critical limitations of each tier.
The comparative advantages of HAPs become evident when evaluated against other network layers across key performance dimensions:

\emph{1) Coverage vs. latency tradeoff}: GEO satellites provide continental-scale coverage but suffer from propagation delays exceeding 250 ms, making them unsuitable for latency-sensitive services \cite{lee2019performance}. LEO satellites reduce latency to 20-50 ms but require large constellations for continuous coverage and introduce frequent handover overhead \cite{abderrahim2020latency, luo2024energy1}. Low-altitude UAVs offer sub-millisecond latency but only cover small areas with hour-scale endurance \cite{luo2024escm}, \cite{zheng2026quantum}, \cite{luo2026real}. HAPs achieve a balanced performance: a single platform can cover a circular area with a 50-200 km radius at an end-to-end latency of 2-10 ms, simultaneously supporting wide-area coverage and real-time service delivery \cite{kurt2021vision, lou2023haps}.

\emph{2) Link budget and deployment cost}: The free-space path loss of HAP-to-ground links is approximately 20-30 dB lower than LEO satellite links and 50-60 dB lower than GEO links at the same frequency \cite{karapantazis2005broadband}, enabling smaller user terminals and lower transmit power. In terms of cost, HAP systems require significantly lower upfront investment than satellite constellations and offer better cost-effectiveness than terrestrial BSs in sparsely populated or geographically challenging regions \cite{svistunov2025bridging, abbasi2024haps}.

\emph{3) Service persistence and deployment flexibility}: Unlike satellites constrained by orbital mechanics, HAPs can be dynamically repositioned to adapt to traffic demand fluctuations, disaster scenarios, or temporary events. Compared to low-altitude UAVs limited by battery endurance, stratospheric HAPs can obtain solar energy and provide uninterrupted service for weeks to months without resupply \cite{mazza2025modeling}. Its continuity is close to that of the satellite system, while maintaining operational flexibility \cite{huang2026high, ccougay2026hap}.

This middle-layer positioning grants HAPs a unique role in connecting upper and lower layers in SAGINs. They can serve as aerial relays to extend satellite coverage to shadowed ground regions, act as edge aggregation points to offload traffic from congested feeder links, and function as local control hubs to coordinate lower-altitude UAV swarms. Without this intermediate layer, SAGINs would face a critical capability gap between high-altitude satellite systems and short-range terrestrial/aerial infrastructure \cite{bakambekova2024interplay}.

\subsection{Functions of HAPs in SAGINs}

Building on their unique middle-layer positioning, HAPs fulfill core functional roles in multi-tier SAGIN ecosystems, covering connectivity provision, service delivery, and network coordination \cite{lou2023haps, huang2026high}.

\subsubsection{Wide-Area Access Node}
As airborne BSs, HAPs provide seamless broadband wireless access to underserved areas including rural regions, deserts, oceans, and disaster zones where terrestrial infrastructure is unavailable or destroyed. 3GPP NTN specifications explicitly identify HAP-based airborne access as a key use case for 5G-Advanced and 6G systems, supporting both direct terminal access and cell-free massive MIMO operation \cite{3gpp38811}. Recent field trials by NTT and AALTO HAPs have demonstrated standard Long-Term Evolution (LTE) connectivity to commercial smartphones via a Zephyr platform, achieving downlink throughput exceeding 4 Mbps over a 20 km stratospheric link \cite{ntt2025haps}.

\subsubsection{Wireless Relay Node}
HAPs act as flexible airborne relays to bridge disconnected network segments and improve link quality \cite{karaman2025demand}. They can relay signals between terrestrial network islands, extend satellite coverage to shadowed ground regions, and provide Non-Line-of-Sight (NLoS) connectivity for mountainous terrain or dense urban canyons. In emergency scenarios, HAP relays can be rapidly deployed to restore communication links within hours when terrestrial infrastructure is damaged \cite{niu2024hap, cao2026survey}. Compared to ground relays, HAP relays offer significantly more extensive coverage and fewer Line-of-Sight (LoS) obstructions.

\subsubsection{Integrated Backhaul Hub}
HAPs serve as aggregation points for access traffic and intermediate nodes for space-ground backhaul \cite{chen2025space}. In Integrated Access and Backhaul (IAB) architectures \cite{cho2025resource}, a single HAP simultaneously provides user access and connects to satellite or terrestrial gateway stations via high-capacity feeder links, eliminating the need for separate backhaul infrastructure \cite{lou2023haps}. For multi-HAP deployments, inter-platform backhaul links can form a meshed aerial backbone network, further enhancing network resilience and reducing dependency on ground gateways \cite{liu2026sum}.

\subsubsection{Aerial Edge Service Node}
Equipped with onboard computing, storage, and caching resources, HAPs function as stratospheric edge servers \cite{hjaiji2026cybersecurity}, \cite{wang2025stratospheric}, bringing service processing closer to end users. They host edge computing tasks for massive IoT devices, proactively cache popular content to reduce backhaul bandwidth consumption, and execute local data processing for onboard and offloaded workloads \cite{ke2021edge, ren2022caching}. For low-altitude UAV swarms and vehicular networks, HAP-hosted edge services reduce round-trip delay by an order of magnitude compared to remote cloud processing, enabling latency-critical applications.

\subsubsection{Low-Altitude Aerial Coordinator}
HAPs act as wide-area control hubs for low-altitude UAV fleets and low-altitude economy infrastructure \cite{erotokritou2025regulatory}. They provide Positioning, Navigation, and Timing (PNT) services for aerial users \cite{mooseli2025evaluation}, coordinate spectrum resource allocation across low-altitude nodes, and perform trajectory planning and collision avoidance for swarm operations \cite{huang2026high}. By centralizing coordination functions at the stratospheric layer, HAPs improve the scalability and reliability of low-altitude networks, especially in regions with dense aerial traffic.

\subsubsection{Cross-Layer Intermediate Orchestrator}
At the system level, HAPs serve as distributed orchestration nodes in multi-layer SAGINs, mediating resource allocation across space, air, and ground segments \cite{toka2025dimensioning}. They dynamically adjust service distribution between satellite backhaul, onboard processing, and terrestrial offloading based on real-time traffic demands, channel conditions, and energy budgets \cite{bakambekova2024interplay, luo2026ai}. This intermediate orchestration capability reduces the signaling overhead of centralized ground core networks and enables end-to-end Quality-of-Service (QoS) guarantees for heterogeneous applications.

\section{Architecture for S$^2$C$^2$I-Integrated HAPs}\label{Sec-III} 
Building on the middle-layer positioning of HAPs in SAGINs, this section establishes a unified S$^2$C$^2$I integration architecture from three dimensions: connectivity planes, payload functional splits, and cross-layer computing-storage continuum. The architecture clarifies the data transmission paths, function deployment boundaries, and hierarchical resource collaboration mechanisms of HAP systems.

\subsection{Connectivity of HAPs in SAGIN}
As the intermediate backbone of SAGINs, HAPs rely on a multi-plane connectivity stack to bridge space, aerial, and ground segments. This stack can be systematically decomposed into five logical link types: service links, HAP-UAV links, feeder links, inter-HAP links, and HAP-satellite links \cite{3gpp38811, lou2023haps, cao2026survey}. Each link serves distinct functional roles and operates at different frequency bands and performance levels, as shown in Table. \ref{tab:link_types}. The layered connectivity structure directly determines the data transmission path, service coverage scope, and resource scheduling granularity of the S$^2$C$^2$I framework.

\begin{table}[!t]
\centering
\caption{Comparison of HAP Connectivity Link Types}
\label{tab:link_types}

\renewcommand{\arrayrulewidth}{0.8pt}
\renewcommand{\tabcolsep}{4pt}

{\fontsize{7.5}{9.5}\selectfont 
\begin{tabular}{m{1.3cm}<{\centering}|m{1.2cm}<{\centering}|m{1.3cm}<{\centering}|m{1.3cm}<{\centering}|m{1.4cm}<{\centering}}
\hline\hline
\rowcolor{gray!15}
\textbf{Link Type} & \textbf{Typical Band} & \textbf{Distance} & \textbf{Typical Rate} & \textbf{SAGIN Plane}\\
\hline
Service Link 
& Sub-6 GHz / mmWave & 20–200 km & 10 Mbps–1 Gbps & Access Plane\\
\hline
HAP-UAV Link
& Sub-6 GHz / mmWave & 5-30 km & 100 Mbps-5 Gbps & Aerial Access Plane\\
\hline
Feeder Link 
& Ka/Ku / E-band / FSO & 20–100 km & 1–50 Gbps & Backhaul Plane\\
\hline
Inter-HAP Link 
& mmWave / FSO & 50–300 km & 1–20 Gbps & Aerial Backbone\\
\hline
HAP-Satellite 
& Ka/Ku-band & 300–1500 km & 100 Mbps–5 Gbps & Space-Air Interface\\
\hline\hline
\end{tabular}}
\vspace{2pt}
\end{table}

\subsubsection{Service Links}
Service links, also known as user links, refer to the wireless connections between HAPs and end users on the ground, low-altitude UAVs, and IoT terminals. As the access interface of the aerial layer, service links directly carry user data traffic and service signaling, corresponding to the access plane of the SAGIN architecture \cite{niu2024hap}.
According to 3GPP NTN specifications, HAP service links can operate in Sub-6 GHz, millimeter Wave (mmWave) \cite{oliveri20246g}, and even optical bands 
\cite{abbasi2024haps}. Sub-6 GHz bands support wide-area coverage with moderate data rates, suitable for massive IoT and rural broadband access \cite{vaezi2022cellular}. mmWave bands provide gigabit-level throughput for high-density user areas, but are more susceptible to atmospheric attenuation \cite{luo2023performance}. Optical bands based on Free-Space Optics (FSO) deliver tens of Gbps capacity with license-free spectrum and electromagnetic interference immunity \cite{wang2024free}, at the cost of higher sensitivity to cloud attenuation and pointing errors \cite{ko2024cloud}. They natively enable sensing-communication convergence by sharing optical apertures with imaging and atmospheric sensing payloads \cite{trinh2025optical}.
Unlike terrestrial BSs, HAP service links enjoy near-LoS propagation at stratospheric altitudes, resulting in more uniform channel quality and larger single-cell coverage radius, typically ranging from 50 km to 200 km \cite{kurt2021vision}. In S$^2$C$^2$I integration, service links not only transmit communication data but also carry sensing control signaling, computing offloading requests, and cached content delivery.

\subsubsection{HAP-UAV Links}
As a dedicated vertical extension of the access plane, HAP-UAV links are Air-to-Air (A2A) wireless connections between stratospheric HAPs and low-altitude UAV fleets, forming a two-tier aerial access hierarchy \cite{huang2026high}. Different from ground-oriented service links affected by terrain shadowing and urban clutter, HAP-UAV links operate in an almost full LoS propagation environment with minimal multipath fading \cite{li2026joint}, supporting stable high-rate transmission even for high-mobility UAV nodes.
Practical deployments usually adopt a dual-band design. First, Sub-6 GHz bands bear flight control signaling, telemetry data, and low-rate sensing streams to guarantee wide coverage and ultra-high reliability. Then, mmWave bands are used for backhauling high-volume raw observation data and offloading computing tasks from UAV swarms to HAP edge servers \cite{liu2024distributed}. The typical slant range of HAP-UAV links ranges from 5 km to 30 km, with throughput scaling from hundreds of Mbps to multiple Gbps depending on frequency band and antenna array configuration \cite{huang2026multi}.
In the S$^2$C$^2$I integration framework, HAP-UAV links serve as the core bearing for layered aerial sensing and collaborative computing. Low-altitude UAVs upload high-resolution perception data and task requests to HAPs via these links, and receive locally processed inference results, cached mission packages, and coordinated trajectory instructions in return \cite{ke2021edge}. 

\subsubsection{Feeder Links and Gateway Connectivity}
Feeder links connect HAPs to ground gateway stations and core networks, undertaking the backhaul of aggregated user traffic, platform control signaling, and management data 
\cite{cho2025resource}. They correspond to the backhaul plane of the SAGIN architecture and are the main bottleneck restricting the throughput of HAP systems. The ground gateway further connects to the terrestrial core network and cloud data centers, forming the end-to-end data path from Air-to-Ground (A2G) \cite{wang2025bridging}.
Feeder links typically use higher frequency bands such as Ka-band, Ku-band, E-band mmWave, and FSO communication to obtain larger bandwidth 
\cite{elkhazraji2025haps, trinh2025optical}. For example, Ka-band feeder links can provide single-link throughput of several Gbps \cite{schwarz2025optical}, while FSO feeder links can reach tens of Gbps. In multi-gateway deployment, HAPs can dynamically switch feeder links between different ground gateways to balance load and avoid link interruption caused by rain attenuation. For S$^2$C$^2$I systems, the capacity of feeder links directly constrains the amount of raw sensing data that can be uploaded to the ground cloud, as well as the update frequency of cached content and AI models \cite{ren2022caching, vallero2022caching}.

\subsubsection{Inter-HAP and HAP-Satellite Links}
Inter-HAP links form a meshed stratospheric backbone network among multiple HAP platforms, enabling horizontal traffic offloading, collaborative sensing, and distributed computing without traversing ground gateways \cite{liu2026sum}. Inter-HAP links usually adopt mmWave or FSO technology with high bandwidth and narrow beam, which can achieve Gbps-level transmission rate between platforms tens to hundreds of kilometers apart \cite{svistunov2025bridging}. In disaster scenarios where ground gateways are unavailable, inter-HAP links can relay traffic to the nearest surviving gateway, greatly improving network resilience.
Moreover, HAP-satellite links connect the aerial layer to the space layer, extending HAP coverage to areas without ground infrastructure \cite{ibrahim2025evaluating}. They are mainly used for global reach backhaul, satellite navigation enhancement, and cross-layer data fusion 
\cite{yu2026topology}. In the S$^2$C$^2$I framework, HAP-satellite links support uploading high-value processed data to satellite cloud storage and downloading global model parameters to HAP edge nodes.

\subsection{Payload Architecture}

The onboard payload architecture of HAPs determines the location of network functions and the maximum S$^2$C$^2$I capability that can be supported. From early bent-pipe relays to fully regenerative base stations, and further to functionally split open architectures, the evolution of HAP payloads reflects the shift from pure connectivity to multi-functional integration \cite{kurt2021vision, svistunov2025bridging}. Four mainstream architectures are summarized in Table \ref{tab:payload_arch_transposed}, with clear boundaries of onboard and ground functions.

\begin{table*}[!t]
\centering
\caption{Comparison of HAP Payload Architectures}
\label{tab:payload_arch_transposed}

\renewcommand{\arrayrulewidth}{0.8pt}
\renewcommand{\tabcolsep}{8pt}

{\fontsize{8}{10}\selectfont 
\begin{tabular}{m{2cm}<{\centering}|m{2cm}<{\centering}|m{1.9cm}<{\centering}|m{1.9cm}<{\centering}|m{2cm}<{\centering}|m{2cm}<{\centering}|m{2.2cm}<{\centering}}
\hline\hline
\rowcolor{gray!15}
\textbf{Architecture} & \textbf{Baseband Processing} & \textbf{Network Functions} & \textbf{Computing} & \textbf{Storage/Caching} & \textbf{End-to-End Latency} & \textbf{Payload Complexity}\\
\hline
Transparent
& Fully on ground & Ground core & None & None & High (double hop) & Very Low\\
\hline
Regenerative
& Fully onboard & Full onboard stack & Full edge computing & Full onboard cache & Low (local processing) & High\\
\hline
CU/DU Split
& DU onboard, CU ground & Distributed deployment & User-plane edge & Content plane onboard & Medium & Medium\\
\hline
Onboard S$^2$C$^2$I
& Sensing pre-processing & Onboard UPF & Local AI inference & Proactive caching & Very Low (closed-loop control) & Very High (cross-function orchestration)\\
\hline
Typical Scenario
& Basic relay coverage & Wide-area access & IoT offloading & Content delivery & Mission-critical services & Multi-mission platforms\\
\hline\hline
\end{tabular}}
\vspace{2pt}

\end{table*}

\subsubsection{Transparent Architecture}
The transparent architecture is the earliest and most mature HAP payload scheme, also known as bent-pipe relaying \cite{lone2026toward}. In this mode, the HAP only performs frequency conversion, amplification, and beamforming on the received Radio Frequency (RF) signal, without any baseband demodulation, decoding, or routing processing \cite{karapantazis2005broadband, 3gpp38811}. All network functions, including baseband processing, core network control, and service scheduling, are completely deployed on the ground gateway station.
The main advantage of the transparent architecture is low payload complexity, low power consumption, and high reliability, which is suitable for platforms with limited payload capacity \cite{guimaraes2025survey}. However, since all signals must be transmitted back to the ground for processing, the end-to-end latency is doubled, and it is impossible to deploy onboard computing, caching, and local sensing processing. Therefore, the transparent architecture can only support basic communication relay functions, and cannot realize the full S$^2$C$^2$I integration capability \cite{lou2023haps}.

\subsubsection{Regenerative Architecture}
The regenerative architecture integrates complete baseband processing units onboard the HAP \cite{kuikel2025triple}, which can independently complete demodulation, decoding, medium access control, and IP layer routing, equivalent to a stratospheric base station \cite{widiawan2007high, abbasi2024haps}. In 3GPP Release 17 and later NTN specifications, regenerative payloads are defined as a key technology to reduce terminal complexity and improve spectrum efficiency \cite{3gpp38811}.
Regenerative HAPs can host a series of onboard network functions, such as Distributed Unit (DU), User Plane Function (UPF), and even local mobility management \cite{veisi2025non}. Thus, this architecture can achieve the S$^2$C$^2$I integration: sensing data can be pre-processed locally without occupying feeder link bandwidth; content and model parameters can be cached onboard to reduce repeated transmission; edge computing tasks can be executed directly on the platform to achieve millisecond-level response \cite{ke2021edge, ren2022caching}. The trade-off is higher payload weight, power consumption, and system complexity.

\subsubsection{CU/DU Split Functional Architecture}
Drawing on the Open-Radio Access Network (O-RAN) functional split paradigm, the research further splits the regenerative base station into Centralized Unit (CU) and DU, forming a layered functional deployment architecture \cite{cumali2026rate}. In this scheme, the DU responsible for physical layer and low-layer protocol processing is deployed on the HAP, while the CU responsible for high-layer Radio Resource Control (RRC) and Packet Data Convergence Protocol (PDCP) protocols can be flexibly placed on the ground gateway or distributed among multiple HAPs.
This architecture balances performance and resource overhead. Low-latency physical layer processing stays onboard to avoid feeder link delay. The resource-intensive control functions are centralized on the ground to reduce payload burden. For S$^2$C$^2$I systems, the CU/DU split supports more flexible function orchestration. That is, computing and caching modules can be co-located with the user plane on the HAP, while global AI model training and network orchestration can be hosted on the ground CU cluster. It is regarded as the most promising architecture for future multi-functional HAP systems.

\subsubsection{Onboard S$^2$C$^2$I Converged Architecture}
Recent studies have integrated partial S$^2$C$^2$I functionalities onto HAP payloads for specific scenarios.
For instance, Ren et al. \cite{ren2022caching} combined communication, edge computing, and content storage on HAPs for intelligent transportation systems, jointly optimizing computation offloading and caching strategies. 
Kanani et al. \cite{kanani2025optimizing} constructed a HAPS-UAV collaborative Integrated Sensing and Communications (ISAC) system where the HAP acts as a central processing unit for sensing data aggregation and computation, covering sensing, communication, and computing. 
Their follow-up work \cite{kanani2025haps} further realized monostatic ISAC on a single HAP via MIMO beamforming, achieving hardware-level communication-sensing co-design while computing and storage still rely on terrestrial support. However, full native integration of all five dimensions remains unexplored in existing works.

Driven by the 6G mission-critical demands such as wide-area disaster response, siloed functional stacking suffers from redundant payload overhead and long cross-module scheduling delays, motivating us to design the onboard S$^2$C$^2$I converged architecture. It natively integrates sensing, storage, communication, computing, and intelligence rather than simple module superposition, and serves as the payload foundation for the unified S$^2$C$^2$I paradigm. In this architecture, the HAP hosts the DU and UPF for communication services, as well as four tightly coupled modules: a sensing pre-processing unit, a hierarchical storage module, an edge AI inference engine, and a cross-function orchestrator. Control-plane functions and global model training remain ground-deployed to conserve onboard resources.
The native coupling of the five functionalities can be characterized by joint resource constraints and closed-loop latency bounds. 
The main tradeoff is the highest system complexity among all architectures, as cross-function coupling introduces joint resource allocation challenges between sensing throughput, storage occupancy, communication rate, computing load, and intelligence accuracy \cite{luo2026best}. This architecture is most suitable for multi-mission HAPs with integrated service demands.

\subsection{Cloud-Edge-HAPs Space Continuum}
\begin{figure}[!t]
\centering
 \includegraphics[width=3.5 in]{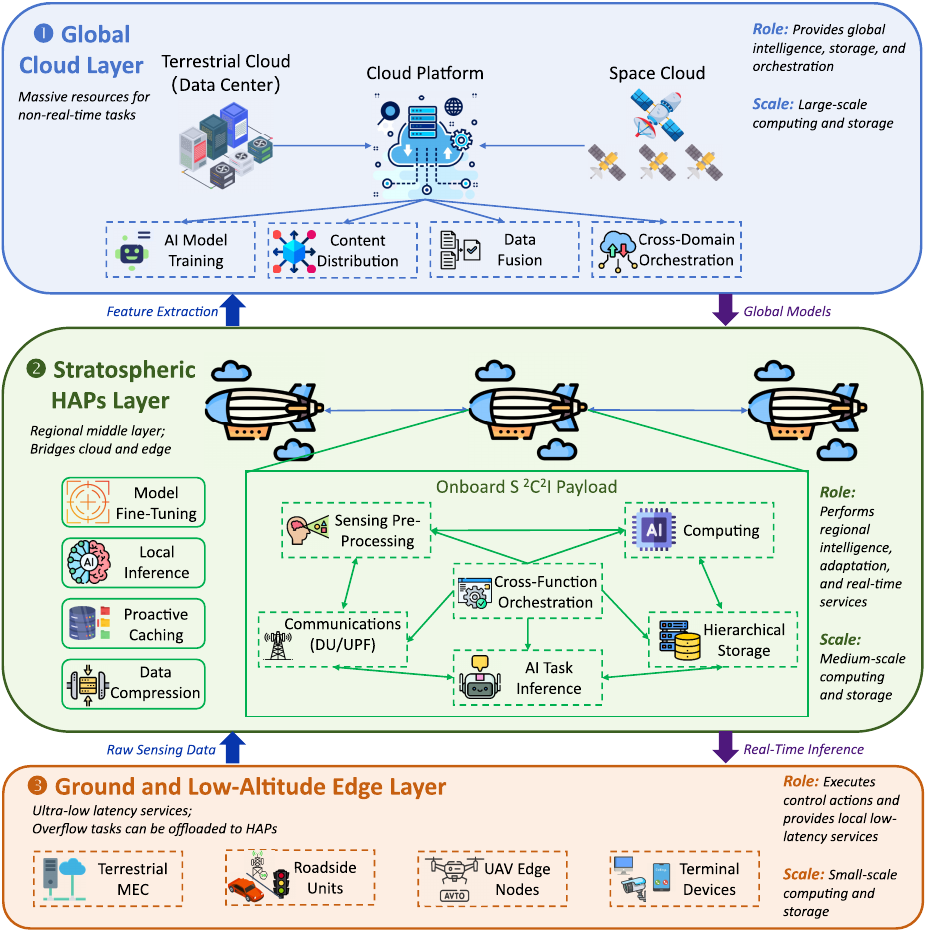}
   \caption{Cloud-Edge-HAPs Space Continuum. It extends S$^2$C$^2$I integration from single platforms to network scale via three hierarchical resource layers and cross-layer functional coupling mechanisms.}
\label{fig:frame}
   \vspace{-0.5cm}
\end{figure}

Building upon the onboard S$^2$C$^2$I converged payload architecture, we propose a cloud-edge-HAPs space continuum. It extends five-dimensional S$^2$C$^2$I integration from single HAPs to SAGINs. 
In this architecture, each satellite equipped with onboard processing capabilities serves as a Mobile Edge Computing (MEC) node \cite{cheng2025scoda}. All satellite nodes can jointly constitute a holistic space cloud with global coverage \cite{kuang2025space}.
As the core hub of the continuum, S$^2$C$^2$I-enabled HAPs bridge the top-tier space cloud and terrestrial cloud infrastructure, mid-tier stratospheric edge nodes, and bottom-tier ground MEC servers and end devices \cite{wang2025bridging}, \cite{ibrahim2025latency}, as shown in Fig. \ref{fig:frame}. It forms a layered resource pool with descending latency and computing scale. By mapping S$^2$C$^2$I functionalities to layers based on latency demands and computing intensity, the proposed architecture achieves an optimal performance-efficiency tradeoff. It also extends the onboard joint resource constraints and closed-loop latency bounds to the network scale.

\subsubsection{Hierarchical Resource Layers}
The continuum consists of three core layers with differentiated S$^2$C$^2$I responsibilities:

\begin{itemize}
    \item \textbf{Global Cloud Layer}: A two-tier global resource pool composed of the network-formed Space Cloud and terrestrial cloud data centers, with aggregated massive computing and storage resources. Each satellite operates as an independent onboard MEC node to perform local on-orbit processing, and realizes collaborative scheduling of cross-satellite computing, storage, and data resources via inter-satellite links \cite{li2025leo}, \cite{sun2025ultra}. This layer is responsible for non-real-time global tasks, including large-scale AI model training, multi-source sensing data fusion, content distribution, and cross-domain orchestration \cite{yuan2025satellite}. It generates base models and scheduling rules pushed to edge nodes, corresponding to the ground-deployed control plane and global training module of the onboard architecture \cite{ge2026space}.
   \item \textbf{Stratospheric HAPs Layer}: Serving as the core regional hub of the entire space continuum, HAPs deployed in the quasi-stationary stratosphere host the full set of onboard S$^2$C$^2$I modules, and fill the performance gap between orbital Space Cloud nodes and terrestrial infrastructure. Compared with the high propagation latency of space cloud and the limited coverage radius of ground MEC servers, HAPs achieve an optimal tradeoff among coverage scale, latency, and onboard computing capacity. 
HAPs can perform regional model fine-tuning, local inference, and proactive caching to avoid repeated feeder link transmission \cite{ren2022caching}, \cite{khan2026multi}.
As distributed on-demand MEC nodes, HAPs directly receive and process computation offloading requests from ground terminals, UAVs, and IoT devices. This will realize localized data processing and service closed-loop without traversing the core network or ISL.
Meanwhile, HAPs upload only high-value feature data to the cloud to reduce backhaul overhead. HAPs also undertake the aggregation, dimensionality reduction, and preprocessing of multi-source data from the bottom edge layer, effectively relieving the computing and transmission pressure of the upper cloud layer.
    \item \textbf{Ground and Low-Altitude Edge Layer}: Terrestrial MEC, Road Side Units (RSUs), and UAV edge nodes delivering millisecond-level services such as autonomous driving control and real-time sensing \cite{zhao2025survey}, \cite{luo2025wireless}. Nodes dynamically offload overflow tasks to the HAP layer when local resources are insufficient.
\end{itemize}

\subsubsection{S$^2$C$^2$I Functional Coupling Mechanism}
Extending the native five-dimensional coupling from single HAP payloads to the network scale, the space continuum realizes cross-layer collaborative synergy of sensing, computing, communication, storage, and intelligence via hierarchical function mapping. 

\begin{itemize}
    \item \textbf{Sensing}: Edge-side raw collection, HAP-side aggregation and dimensionality reduction, and cloud-side multi-source fusion and archival storage. Sensing sampling and upload strategies are dynamically tuned by intelligence and storage status \cite{zhu2025survey}, with HAPs converting heterogeneous edge data into structured features for upper-layer fusion.
     \item \textbf{Storage}: Massive cold archival storage in the cloud, regional hot storage, and proactive caching at HAPs, and lightweight real-time storage at the ground edge. As the hot-cold data conversion hub, HAPs reduce redundant cross-layer transmission and improve global storage utilization via intelligent cache prefetching \cite{wang2026survey}.
    \item \textbf{Communication}: The underlying carrier for all cross-layer interactions, covering links described in Section \ref{Sec-III}-A. Intelligence-driven dynamic bandwidth and routing adaptation match communication resources with sensing, computing, and storage demands \cite{luo2026trustworthy}. HAPs act as regional relay nodes to bridge wide-coverage satellite links and high-density ground access \cite{he2024nonterrestrial}.
    
    \item \textbf{Computing}: The global cloud undertakes large-scale model training, the HAPs layer acts as a regional computing relay for model fine-tuning, and the ground edge handles lightweight tasks. Intelligent cross-layer task offloading realizes collaborative computing-storage scheduling, with HAPs balancing efficiency and transmission overhead by undertaking overflow tasks from the bottom and sharing pressure from the top \cite{heidari2023machine}.
   
    \item \textbf{Intelligence}: It drives collaborative optimization of the others, including sensing filtering, computing scheduling, communication resource allocation, and storage lifecycle management \cite{luo2026ai}. HAPs support independent regional closed-loop intelligence while feeding incremental optimization back to the cloud for global iteration.
\end{itemize}

This network-scale five-dimensional coupling mechanism breaks the siloed deployment of traditional independent functional systems. Anchored by the HAPs layer, it enables cross-layer coordinated scheduling of multi-dimensional resources, 
and improving overall resource utilization and end-to-end service performance.

\subsubsection{Cross-Layer Data and Control Loops}
Complementing the local closed loop within individual HAP payloads, two cross-layer flows enable network-wide S$^2$C$^2$I coordination:

\begin{itemize}
    \item \textbf{Data processing loop}: Raw sensing data is pre-processed at the ground/UAV layer, further dimension-reduced at the HAP layer via onboard pre-processing modules. Only high-value data can be uploaded to the cloud for long-term storage \cite{zhu2025optimized}. This layered compression cuts cross-layer data volume by orders of magnitude and alleviates onboard resource constraints.
    \item \textbf{Control and model loop}: The cloud trains global models and distributes parameters to HAPs. They perform local adaptation and real-time inference before issuing control instructions to the bottom layer. Bottom-layer feedback is uploaded incrementally to iteratively optimize upper-layer models \cite{qian2022distributed}. This forms a hierarchical closed-loop system with the onboard local loop: the onboard loop guarantees the end-to-end latency bound for mission-critical services, while the cross-layer loop improves long-term model accuracy and global resource efficiency.
\end{itemize}

The proposed continuum fully exploits HAPs’ middle-layer advantage to offset the propagation latency of the orbital Space Cloud and remote terrestrial clouds, as well as the limited computing power of the bottom-edge nodes. It provides a unified framework for full-stack S$^2$C$^2$I deployment in SAGINs and forms a coherent logic from single-platform payload design to network-wide resource orchestration.

\section{Enabling Technologies for S$^2$C$^2$I-Integrated HAPs}\label{Sec-IV} 

\subsection{RF, mmWave, THz, FSO, and Hybrid Connectivity}

Connectivity provides the physical foundation for exchanging user traffic, sensing observations, cached content, computing tasks, AI models, and control information in S$^2$C$^2$I-integrated HAP networks. Since these data flows differ considerably in rate, reliability, delay, and energy requirements, no single transmission medium can efficiently support all of them. 
Their representative designs and S$^2$C$^2$I roles are summarized in Table~\ref{tab:multi_medium_haps}.

\begin{table*}[!t]
\centering
\caption{Representative HAP Communication Technologies and Their S$^2$C$^2$I Roles}
\label{tab:multi_medium_haps}
\renewcommand{\arrayrulewidth}{0.8pt}
\renewcommand{\tabcolsep}{3.2pt}
{\fontsize{7.1}{8.6}\selectfont
\begin{tabular}{m{0.75cm}<{\centering}||m{0.65cm}<{\centering}|m{2.3cm}<{\centering}|m{2.4cm}<{\centering}|m{3cm}<{\centering}|m{7.4cm}<{\centering}}
\hline \hline
\rowcolor{gray!15}
\textbf{Ref.} & \textbf{Year} & \textbf{Medium} & \textbf{Link/Topology} & \textbf{S$^2$C$^2$I Role} & \textbf{Representative Insight}\\
\hline
\rowcolor{blue!10}
\cite{shibata2020system} & 2020 & Sub-6-GHz RF & HAPs-UE multi-cell service & Wide-area access and energy-aware communication & Multi-layer cellularization enables gigabit-class service and can reduce total transmit power for a fixed footprint\\
\hline
\rowcolor{blue!10}
\cite{hsieh2020uav} & 2020 & Sub-6 RF + mmWave & Service and feeder links & Access/backhaul separation & Sub-6 service supports broad coverage; a 28-GHz feeder carries aggregate multi-cell traffic with high availability\\
\hline
\rowcolor{blue!10}
\cite{odeyemi2022reconfigurable} & 2022 & RF & Source-RIS-HAPs-users & Programmable propagation and robust access & Increasing RIS elements improves the relayed multiuser link relative to a non-RIS baseline.\\
\hline
\rowcolor{blue!10}
\cite{salehi2023reliability} & 2024 & RF & SAGIN paths & Mission-critical control loop & Multi-connectivity remains necessary even with terrestrial interference mitigation and practical backhaul failures\\
\hline
\rowcolor{blue!10}
\cite{lin2025connectivity} & 2025 & RF & H2D, H2G2D, and hybrid coverage & Large-scale IoT/sensing continuity & Critical HAPs/gateway densities identify the transition to large-scale connected service regions\\
\hline
\rowcolor{orange!12}
\cite{dutta2019haps} & 2019 & 28-GHz mmWave & HAPs-users service & High-rate access with control-plane adaptation & PRACH, rather than link budget alone, limits large footprints; the proposed procedure approximately doubles usable area\\
\hline
\rowcolor{orange!12}
\cite{grieco2023preliminary} & 2023 & 20-100-GHz RF & GEO-HAPs & Space--air backhaul & High-rate operation is feasible in selected bands, while oxygen absorption causes a marked loss near 60~GHz\\
\hline
\rowcolor{orange!12}
\cite{kanani2025optimizing} & 2026 & 120-GHz sub-THz + 2.4-GHz RF & HAPs-UAV backhaul and UAV-users access & Sensing-communication-computing integration & The HAPs centralizes processing and coordination, while backhaul carries high-volume sensing and communication data\\
\hline
\rowcolor{green!10}
\cite{ata2022haps} & 2023 & FSO & Uplink, downlink, inter-HAP & Optical backhaul and sensing support & Downlink is generally stronger than uplink; adaptive optics mitigates turbulence but exhibits diminishing returns.\\
\hline
\rowcolor{green!10}
\cite{zedini2024improving} & 2024 & MIMO FSO & Ground--HAPs and inter-HAP & High-volume sensing/model transport & Spatial diversity counters turbulence, pointing error, and attenuation\\
\hline
\rowcolor{purple!10}
\cite{shah2021adaptive} & 2021 & Hybrid FSO/RF & Ground-HAPs-satellite & Weather-resilient bulk-data path & Adaptive combining improves on hard switching; the HAPs relay is especially useful for uplink reliability\\
\hline
\rowcolor{purple!10}
\cite{swaminathan2021haps} & 2021 & Hybrid FSO/RF & Ground-HAPs-satellite & Reliable SAGIN feeder & RF backup gives larger gains as pointing, zenith-angle, and wind impairments become more severe\\
\hline
\rowcolor{purple!10}
\cite{deka2022performance} & 2022 & Hybrid FSO/RF & Ground-HAPs-ground & Energy-aware relay operation & Optimized source/HAPs power improves performance by more than 3~dB over equal allocation in the studied case\\
\hline
\rowcolor{purple!10}
\cite{liu2023joint} & 2024 & FSO backhaul + RF access & Satellite-HAPs-ground & Communication-computing resource coupling & Backhaul quality, payload limits, interference, and access decisions should be optimized jointly\\
\hline
\rowcolor{purple!10}
\cite{ali2024optimizing} & 2024 & FSO backhaul + RF broadcast & Satellite-HAPs-ground & Intelligent cross-layer orchestration & Learning accelerates power control and scales to multiple HAPs and devices while approaching model-based performance\\
\hline \hline
\end{tabular}}
\vspace{2pt}
\end{table*}

\subsubsection{RF Connectivity}

Sub-6-GHz and microwave RF remain the primary technologies for HAP service links because of their broad coverage, relatively relaxed alignment requirements, and compatibility with terrestrial mobile terminals \cite{tezergil2022wireless}. Shibata et al. optimized HAPs cellular layouts and showed that the number, arrangement, and antenna patterns of cells should be jointly designed according to the required footprint \cite{shibata2019study}. Their subsequent work demonstrated that multi-cell frequency reuse can support gigabit-class aggregate capacity while reducing the transmission power through appropriate cell and antenna optimization \cite{shibata2020system}. Similarly, the New Radio (NR) architecture in \cite{hsieh2020uav} employs adaptive beam steering to maintain fixed ground cells during HAPs movement, using sub-6-GHz RF for user access and mmWave for feeder transmission.
RF connectivity can also be enhanced through programmable propagation and multi-connectivity \cite{qamar2024multi}. Reconfigurable Intelligent Surface (RIS)-assisted HAPs relaying improves outage, bit-error rate, and capacity when direct links are blocked \cite{odeyemi2022reconfigurable}, \cite{ye2022nonterrestrial}. For safety-critical aerial communications, combining terrestrial, A2A, HAPs, and satellite links is more effective than relying on a single access path, particularly under practical antenna patterns and unreliable backhaul \cite{salehi2023reliability}. From a large-scale perspective, percolation analysis identifies the HAPs and gateway densities required to establish continuous HAP-to-device or gateway-assisted coverage \cite{lin2025connectivity}. Integrated HAPs-LEO-terrestrial architectures further show that HAPs can complement sparse terrestrial infrastructure and improve IoT connectivity in remote and disaster-affected areas \cite{de2026integrating}.
These characteristics make RF the reliability and control plane of S$^2$C$^2$I-integrated HAPs. It is particularly suitable for command signaling, emergency traffic, low-rate sensing, multicast delivery, and direct access by commodity terminals. However, because its spectrum is limited, sensing, storage, and computing functions should filter, compress, or aggregate high-volume data before RF transmission.

\subsubsection{mmWave and Sub-THz Connectivity}

mmWave communication provides much larger bandwidth than sub-6-GHz RF \cite{luo2024energy}, but long HAPs slant ranges, and narrow beams introduce power, synchronization, and access challenges. In \cite{dutta2019haps}, a 28-GHz HAPs system achieves significantly higher rates than conventional cellular links, while its coverage is constrained by the NR physical random-access procedure rather than only by the link budget. A retransmission-based Physical Random Access Channel (PRACH) mechanism is therefore introduced to extend the service area. Meanwhile, high-frequency radio is also attractive for feeder and space-air links. For instance, the 28-GHz feeder design in \cite{hsieh2020uav} supports the aggregate traffic of multiple HAPs cells under the considered availability requirement. A GEO-to-HAPs study over 20-100 GHz shows that the achievable Signal-to-Noise ratio (SNR) strongly depends on carrier frequency and atmospheric absorption, with a pronounced degradation around the oxygen-absorption region near 60 GHz \cite{grieco2023preliminary}. Hence, carrier selection must account for atmospheric windows rather than bandwidth alone.
Moreover, sub-THz communication can further support sensing-intensive aerial networks. In \cite{kanani2025optimizing}, a 120-GHz HAPs-UAV backhaul transports communication and sensing information to a HAP acting as a central processing unit, while UAVs serve ground users over lower-frequency RF links.
This architecture closely reflects S$^2$C$^2$I integration: high-capacity links deliver sensing observations; onboard computing performs centralized processing; and intelligent beamforming jointly optimizes communication and sensing performance. 
Nevertheless, beam tracking, absorption loss, phase noise, and payload power consumption \cite{luo2023performance1} imply that mmWave and sub-THz links are better used as high-capacity data planes supported by a more robust RF control plane.

\subsubsection{FSO Connectivity}

FSO communication is well-suited to HAPs feeder, inter-HAP, and satellite-HAPs links because of its large unlicensed bandwidth, high directivity, and low electromagnetic interference \cite{tarhouni2025free}. Its performance, however, is sensitive to turbulence, attenuation, pointing errors, angle-of-arrival fluctuations, clouds, and platform vibration.
A unified analysis of ground-to-HAPs, HAPs-to-ground, and inter-HAPs optical links shows that downlinks generally outperform uplinks and that adaptive optics can effectively mitigate turbulence, although its improvement eventually saturates as residual impairments dominate \cite{ata2022haps}. Additionally, Multiple-Input Multiple-Output (MIMO) optical transmission provides additional spatial diversity against turbulence, pointing errors, and angle-of-arrival fluctuations, thereby improving outage probability, bit-error rate, and ergodic capacity \cite{zedini2024improving}.
Within S$^2$C$^2$I, FSO is particularly suitable for transmitting high-resolution sensing data, digital maps, cached contents, and large AI-model updates. Its intermittent availability also creates a direct coupling with storage and computing \cite{karmous2024can}: data can be buffered during optical blockage, locally compressed or processed, and transmitted when atmospheric and alignment conditions improve.

\subsubsection{Hybrid RF/FSO Connectivity}

Hybrid RF/FSO communication combines the capacity of optical links with the robustness of RF links. Existing studies mainly adopt switching, combining, heterogeneous per-hop transmission, or cross-layer resource optimization.
Threshold-based hybrid HAPs relaying can substantially reduce the error probability of satellite downlinks when turbulence, wind, or pointing errors degrade the optical channel \cite{swaminathan2020performance}. Instead of completely switching between media, the adaptive-combining scheme in \cite{shah2021adaptive} continuously uses FSO and activates the RF link when optical quality deteriorates. Thus, it can achieve better reliability than single-link FSO and conventional hard switching.
Additionally, different media can also be assigned to different network segments. In the uplink SAGIN of \cite{swaminathan2020performance}, a hybrid RF/FSO ground-HAPs hop is combined with an optical HAPs-satellite hop, with RF backup becoming increasingly valuable under severe optical impairments. For ground-HAPs-ground communication, selection combining and joint source/HAPS power allocation improve outage and error performance compared with equal power allocation \cite{deka2022performance}. In multicast networks, the HAPs-assisted RF/FSO/RF architecture provides higher energy efficiency, whereas the longer satellite-assisted RF/FSO/FSO/RF route offers better outage performance and wider reach \cite{yahia2022haps}.
Recent studies further integrate link selection with network optimization. Liu et al. \cite{liu2023joint} jointly optimize user association and beamforming in a satellite-HAPs-ground network using FSO backhaul and RF access. Similarly, \cite{ali2024optimizing} optimizes power allocation for FSO LEO-HAPs links and RF HAPs-device links through both model-based and reinforcement-learning methods. 
These works demonstrate that hybrid connectivity should not be treated merely as physical-layer backup; it should be coordinated with user association, power allocation, computing load, and traffic priority.
\begin{lessonslearned}

The reviewed studies indicate that S$^2$C$^2$I-integrated HAPs should employ heterogeneous links according to the value and urgency of the carried information. 
More importantly, communication decisions should be coordinated with the remaining S$^2$C$^2$I functions. When feeder capacity decreases, onboard computing can extract features or complete inference locally, and storage can buffer delay-tolerant sensing data. When high-capacity links recover, cached observations and model updates can be transmitted in batches. Intelligence can predict traffic, atmospheric conditions, platform motion, energy state, and task deadlines to jointly select the transmission medium, route, beam, coding rate, and processing location. 
\end{lessonslearned}

\subsection{Sensing Payloads and ISAC for HAPs}

Sensing transforms a HAP from a communication relay into a persistent observation and decision node. Its payload may include radar or SAR, optical/environmental sensors, and dual-functional radio arrays that reuse spectrum, waveforms, and RF hardware for sensing and communication. In an S$^2$C$^2$I-integrated HAP, sensing data are further processed, stored, exchanged, and exploited for tracking, beam control, resource allocation, and mission adaptation. Representative HAP-oriented ISAC designs are summarized in Table~\ref{tab:sensing_isac_haps}.

\begin{table*}[!t]
\centering
\caption{Representative HAP Sensing and ISAC Designs and Their S$^2$C$^2$I Roles}
\label{tab:sensing_isac_haps}
\renewcommand{\arrayrulewidth}{0.8pt}
\renewcommand{\tabcolsep}{3.0pt}
{\fontsize{7.1}{8.6}\selectfont
\begin{tabular}{m{0.75cm}<{\centering}||m{0.65cm}<{\centering}|m{2.3cm}<{\centering}|m{2.4cm}<{\centering}|m{3cm}<{\centering}|m{7.4cm}<{\centering}}
\hline \hline
\rowcolor{gray!15}
\textbf{Ref.} & \textbf{Year} & \textbf{Payload/ISAC Mode} & \textbf{Link/Topology} & \textbf{S$^2$C$^2$I Role} & \textbf{Representative Insight}\\
\hline
\rowcolor{blue!10}
\cite{kanani2025haps} & 2025 & Monostatic MIMO/MISO ISAC & HAPs-users/targets & Wide-area sensing and access & Max-min beampattern optimization improves worst-direction sensing while satisfying user SINR and power constraints\\
\hline
\rowcolor{blue!10}
\cite{kanani2025max} & 2025 & Fairness-aware beamforming & HAPs-targets & Fair resource allocation & Joint beamforming equalizes target illumination and minimum user performance under HAPs power limits\\
\hline
\rowcolor{blue!10}
\cite{kanani2025haps1} & 2025 & Dual-role HAPs-ISAC & HAPs and HAPs-UAV hierarchy & Regional sensing and control & HAPs alternate between a wide-area ISAC base station and a CPU for cooperative UAVs\\
\hline
\rowcolor{orange!12}
\cite{zhang2026design} & 2026 & SAR and beamforming & Quasi-stationary/dynamic HAPs & Joint imaging and flight control & Joint 3-D placement or trajectory and beamforming balances communication rate, SAR gain, SNR, and energy\\
\hline
\rowcolor{orange!12}
\cite{huang2025design} & 2025 & Integrated waveform & HAP SAR-ground user & Waveform-level integration & Frequency-index and QAM symbols increase data rate, while phase compensation preserves coherent SAR processing\\
\hline
\rowcolor{orange!12}
\cite{kirik2025isac} & 2025 & ISAC-assisted beam alignment & Inter-HAP Network & Sensing-assisted backhaul & Low-frequency coarse sensing narrows high-frequency beam search and sharply reduces alignment overhead\\
\hline
\rowcolor{green!10}
\cite{kanani2025optimizing} & 2026 & Distributed UAV ISAC with HAPs & UAV-users/targets-HAPs & Sensing aggregation and offloading & UAVs collect echoes and serve users, while HAPs coordinate power allocation and processing\\
\hline
\rowcolor{green!10}
\cite{zhou2025trade} & 2026 & Multi-UAV ISAC with HAP MEC & UAV sensing-HAP & Energy-aware orchestration & Joint scheduling, power, and trajectory optimization balances sensing data acquisition and total energy\\
\hline
\rowcolor{purple!10}
\cite{benaya2025aerial} & 2025 & Full-duplex secure ISAC & HAPS, edge server, and jammer UAV & Secure sensing and computing & Joint beamforming and jammer trajectory satisfy radar, security, offloading, latency, and power constraints\\
\hline
\rowcolor{purple!10}
\cite{ciloglu2026isac} & 2026 & Doppler-assisted disaster ISAC & HAPs backhaul plus UAV BSs & Resilient access and awareness & Shared signals support both communication and Doppler-based detection of victims and mobile users\\
\hline \hline
\end{tabular}}
\vspace{2pt}
\end{table*}

\subsubsection{HAP-Borne Radar and SAR Payloads}

The wide footprint and long endurance of HAPs make radar and SAR attractive for persistent surveillance, environmental monitoring, and disaster assessment. Unlike independent sensing payloads, integrated SAR and communication systems must share antenna aperture, power, waveform dimensions, and flight time. Zhang et al. jointly optimize the 3-D placement or trajectory of quasi-stationary or dynamic HAPs with communication and SAR beamforming \cite{zhang2026design}. The optimized platform location lies between the user and imaging regions, showing that platform motion and synthetic-aperture formation should be co-designed with communication.
At the waveform level, Huang et al. \cite{huang2025design} divide each pulse into frequency-hopping sub-pulses and embed frequency-index and Quadrature Amplitude Modulation (QAM) symbols. Phase compensation restores coherent SAR processing, while a two-step receiver detects the communication symbols. More sub-pulses increase data carriage but degrade range resolution, Bit Error Rate (BER), and processing complexity. Hence, an S$^2$C$^2$I controller should adapt waveform granularity to mission needs, process echoes onboard, and forward only images, detections, or selected regions when feeder capacity is limited.

\subsubsection{MIMO Beamforming and Sensing-Assisted Links}

Large HAPs arrays can illuminate multiple sensing directions while serving users. Kanani et al. \cite{kanani2025max} maximize the minimum sensing beampattern gain under user-SINR and power constraints for MISO/MIMO users over Rician channels \cite{kanani2025haps}. Their fairness-oriented design avoids sacrificing weak sensing directions and outperforms UAV-based ISAC in coverage, sensing gain, and resource fairness. A related max--min formulation further equalizes target illumination and user rates, although complexity increases with antennas, users, and targets. The roadmap in \cite{kanani2025haps1} therefore highlights scalable learning, robust channel modeling, and real-time control as key deployment requirements.
Sensing can also directly support aerial links. Kirik et al. \cite{kirik2025isac} use low-frequency sensing and triggering signals for coarse inter-HAP alignment, followed by high-frequency fine search within a reduced angular region. This method approaches exhaustive-search performance with much lower alignment delay, making it suitable for S$^2$C$^2$I aerial backbones carrying sensing products, cached data, and computing results. Fan et al. \cite{fan2025mobility} similarly use echo-based extended Kalman filtering to predict beams for a mobile energy receiver in sensing-assisted Simultaneous Wireless Information and Power Transfer (SWIPT). The feedback-reduction principle can also be extended to stratospheric platforms serving mobile sensors or UAVs.

\subsubsection{Cooperative HAPS--UAV Sensing and Computing}

In hierarchical architectures, UAVs provide flexible local sensing, while HAPs offer persistent coverage, backhaul, and regional processing. Kanani et al. \cite{kanani2025optimizing} let UAVs transmit joint sensing/communication signals, collect echoes, and relay information to a HAP's CPU over a sub-THz backhaul. Power allocation jointly balances the minimum user SINR and target echo power. This design shows that sensing quality depends not only on the air-to-ground link but also on backhaul capacity, synchronization, and processing delay.
Zhou and Liu study a multi-UAV ISAC system in which UAVs sense ground users and offload part of the workload to a HAP MEC server \cite{zhou2025trade}. Joint optimization of sensing scheduling, transmit power, and trajectories reduces energy and increases collected sensing data compared with ground-processing and fixed-trajectory baselines. Thus, an S$^2$C$^2$I scheduler should jointly decide what to sense, where to process it, and whether to retain raw measurements or transmit extracted features according to deadline, channel quality, CPU load, storage occupancy, and propulsion energy.

\subsubsection{Secure and Mission-Aware ISAC}

Shared sensing and communication signals also introduce security and privacy risks. Benaya et al.  \cite{benaya2025aerial} consider a full-duplex HAPs-ISAC base station that serves users, senses targets, offloads data to an edge server, and guides a jammer UAV toward a detected eavesdropper. Joint transmit/receive beamforming and jammer-trajectory optimization couple sensing, communication, computing, and physical-layer security in a closed control loop.
For disaster response, Ciloglu et al. \cite{ciloglu2026isac} use HAPs as wide-area backhaul and coordination infrastructure for multiple UAV base stations. The same transmissions support user access and Doppler-based detection of stationary or mobile victims, achieving high motion-detection sensitivity and accuracy under terrestrial network failures. In such cases, the sensing objective should be mission utility rather than beampattern gain alone. Detected people, hazards, and damaged infrastructure must be prioritized according to urgency and confidence.
\begin{lessonslearned}
HAP sensing payloads should be designed as part of an end-to-end S$^2$C$^2$I loop rather than as isolated instruments. Waveforms, beams, platform trajectories, UAV cooperation, computing placement, storage, and security response are tightly coupled. Future designs should therefore optimize estimation or imaging fidelity together with latency, energy, data freshness, processing cost, and mission value. 
\end{lessonslearned}

\subsection{On-Board Computing and Stratospheric Edge System}

On-board computing transforms a HAP from a connectivity node into a persistent stratospheric edge system. It can execute offloaded workloads, aggregate sensing streams, host application data and AI models, and coordinate resources across the space-air-ground continuum. Unlike terrestrial MEC servers, however, a HAP must provide these services under strict payload, power, thermal, backhaul, and endurance constraints. Computing is therefore tightly coupled with the other S$^2$C$^2$I functions: sensing determines the generated workload; storage supplies reusable data and buffers intermediate results; communication determines feasible offloading paths; and intelligence predicts demand and orchestrates processing placement. Representative HAP-oriented computing designs are summarized in Table~\ref{tab:onboard_computing_haps}.

\begin{table*}[!t]
\centering
\caption{Representative HAP On-Board Computing and Stratospheric Edge Designs and Their S$^2$C$^2$I Roles}
\label{tab:onboard_computing_haps}
\renewcommand{\arrayrulewidth}{0.8pt}
\renewcommand{\tabcolsep}{3.0pt}
{\fontsize{7.1}{8.6}\selectfont
\begin{tabular}{m{0.75cm}<{\centering}||m{0.65cm}<{\centering}|m{2.3cm}<{\centering}|m{2.4cm}<{\centering}|m{3cm}<{\centering}|m{7.4cm}<{\centering}}
\hline \hline
\rowcolor{gray!15}
\textbf{Ref.} & \textbf{Year} & \textbf{Computing Paradigm} & \textbf{Link/Topology} & \textbf{S$^2$C$^2$I Role} & \textbf{Representative Insight}\\
\hline
\rowcolor{blue!10}
\cite{ren2022caching} & 2022 & Three-tier HAPs edge with caching & Vehicles-RSUs-HAPs & Storage-communication-computing coordination & Joint caching, offloading, bandwidth, and CPU allocation reduces both data-retrieval and execution delay\\
\hline
\rowcolor{blue!10}
\cite{traspadini2023real} & 2023 & Queue-aware HAP edge computing & Ground vehicles-HAP & Real-time sensing-data processing & Sufficient HAP capacity makes real-time object-processing feasible in rural areas\\
\hline
\rowcolor{blue!10}
\cite{nguyen2023hap} & 2023 & RSMA-enabled HAP& Vehicle users-HAP & Communication-computing control & DDPG jointly selects offloading ratio, RSMA split, transmit power, and decoding order to improve task success\\
\hline
\rowcolor{blue!10}
\cite{traspadini2022uav} & 2022 & UAV/HAP-assisted edge continuum & Vehicles-UAVs/HAP & Multi-tier processing placement & Queueing analysis shows that HAP assistance lowers processing time and that hybrid UAV-HAP execution further improves latency\\
\hline
\rowcolor{green!10}
\cite{jia2022hierarchical} & 2023 & Hierarchical aerial MEC & IoT devices-UAVs-HAP & Regional task aggregation and computing & UAV-HAP hierarchy to compute more data and serve more IoT devices than either tier alone\\
\hline
\rowcolor{green!10}
\cite{nabi2025joint} & 2025 & Binary/partial hierarchical offloading & Ground users-UAVs-HAP & Load-balanced edge orchestration & Reduce delay and energy while improving load balance and task-completion ratio\\
\hline
\rowcolor{green!10}
\cite{li2026joint} & 2026 & Multi-UAV/HAP computing & Ground users-UAVs-HAP & Mode and ratio adaptation & A DDQN--PPO decomposition selects discrete offloading modes and continuous ratios, reducing weighted latency and energy\\
\hline
\rowcolor{green!10}
\cite{liu2025energy} & 2025 & UAV/HAP-assisted IoT MEC & IoT-UAVs/HAP & Queue-stable intelligent computing & Lyapunov optimization and hybrid DRL jointly control offloading and resources, minimizing long-term energy\\
\hline
\rowcolor{orange!12}
\cite{mei2022energy} & 2022 & HAP--satellite collaborative MEC & Ground users-HAP-LEO & Energy-aware space-air computing & Joint task assignment and communication/computing-resource allocation reduces weighted energy\\
\hline
\rowcolor{orange!12}
\cite{wu2025multi} & 2025 & Multi-HAP-assisted maritime MEC & Sea users-HAPs-LEO & SAGIN computation continuum & Graph matching, CPU allocation, and alternating optimization coordinate task-HAP and task-satellite association to reduce end-to-end delay\\
\hline
\rowcolor{orange!12}
\cite{yang2025task} & 2025 & SAGIN edge computing & Ground-HAPs/LEOs and HAP-HAP/LEO & Prediction-driven cross-layer orchestration & Task-demand prediction guides MAPPO to avoid overload and resource waste while improving energy efficiency and CPU/bandwidth utilization\\
\hline
\rowcolor{purple!10}
\cite{abderrahim2023leverage} & 2023 & Data center-enabled HAP & Terrestrial data center-HAPs & Solar-powered green computing & One DC-HAP saves about 12\% of electricity cost and a four-HAP deployment about 36\% in the studied setting\\
\hline
\rowcolor{purple!10}
\cite{lu2025distributed} & 2025 & Distributed multi-HAP cloud & Ground-multiple DC-HAPs & Heterogeneous workload migration & A feasibility-preserving formulation and tailored solvers expose the performance, complexity trade-off of large multi-platform computing systems\\
\hline
\rowcolor{purple!10}
\cite{abderrahim2025green} & 2025 & Quantum-computing-enabled HAP & QDCs--FSO-HAP/satellite & Green emerging computing & At the favorable stratospheric altitude, the studied QC-HAP reduces energy use by 21\% and supports 30\% more qubits than a terrestrial QDC\\
\hline
\hline \hline
\end{tabular}}
\vspace{2pt}
\end{table*}

\subsubsection{HAP-Centric Edge Computing for Terrestrial Access}

HAP computing is particularly attractive where vehicles and IoT devices generate latency-sensitive sensing workloads but terrestrial edge infrastructure is sparse. Ren et al. \cite{ren2022caching} integrate vehicles, terrestrial edge nodes, and a HAPs data library, with RSUs caching application fundamentals such as maps and environmental information. Their joint design shows that computing and storage should not be optimized separately. Caching removes repeated HAPs retrieval, whereas HAPs computing supplements overloaded or weak terrestrial servers. For real-time vehicular processing, Traspadini et al. \cite{traspadini2023real} model both ground vehicles and the HAP as queues and derive the partial-offloading factor that maximizes the probability of meeting a frame deadline. Their results indicate that neither fully local nor indiscriminate full offloading is generally optimal because radio transmission, HAP execution, and queueing delays compete. In the studied setting, real-time service at 10 frames/s requires more than 3000 GFLOPS at the HAP.
Communication design also determines usable computing capacity. Nguyen and Park combine HAP with Rate-Splitting Multiple Access (RSMA) to control offloading jointly, message splitting, transmit power, and decoding order \cite{nguyen2023hap}. Traspadini et al.  \cite{traspadini2022uav} further compare local, UAV, HAP, and hybrid aerial execution, showing that a HAP can absorb computationally heavy workloads while UAVs provide a shorter access path. In an S$^2$C$^2$I system, the HAP should cache maps and models; process vehicle sensor streams; and return compact detections or control decisions rather than repeatedly transporting raw data through the feeder link.

\subsubsection{Hierarchical HAP-UAV Collaborative Computing}

A hierarchical aerial edge separates local access from regional processing. Jia et al. \cite{jia2022hierarchical} have UAVs receive IoT tasks and either execute them locally or relay them to a HAP. It can use matching with externality elimination and heuristic adjustment to maximize the amount of data successfully computed. The cooperative hierarchy outperforms HAP-only and UAV-only modes because UAV proximity reduces access cost while the HAP contributes persistent, high-capacity processing. Nabi and Moh \cite{nabi2025joint} extend this principle through binary ground-user-to-UAV offloading and partial UAV-to-HAP offloading. Matching handles association, while enhanced soft actor-critic controls task partitioning and CPU allocation.
More dynamic designs use learning to separate discrete and continuous decisions. Li et al. \cite{li2026joint} employ Double Deep Q-Network (DDQN) \cite{chi2024task} for offloading-mode selection and Proximal Policy Optimization (PPO) \cite{zhang2023towards} for offloading-ratio control in a cooperative multi-UAV/HAP network. Liu et al. \cite{liu2025energy} combine Lyapunov optimization \cite{tong2022dynamic} with hybrid Deep Reinforcement Learning (DRL) so that an UAV/HAP-assisted IoT system can minimize energy without destabilizing long-term task queues. These studies imply that the HAP should operate as a regional compute and intelligence hub. UAVs collect sensing data and offer low-latency access, whereas the HAP stores shared models; processes aggregated workloads; and coordinates association, routing, and resource allocation according to channel, queue, energy, and mission states.

\subsubsection{Space-Air-Ground Cooperative Edge Continuum}

HAPs can also bridge terrestrial and even maritime devices and orbital computing resources. Mei et al.  \cite{mei2022energy} study a HAP cooperating with multiple LEO satellites and jointly optimize task placement and communication/computing resources to reduce weighted energy consumption. Wu et al.  \cite{wu2025multi} consider maritime users in a multi-HAP-assisted SAGIN, where tasks are associated with HAPs and may be forwarded to a LEO satellite. In this hierarchy, HAPs provide a lower-latency first processing tier, while satellites offer overflow capacity and broader reach.
Static decisions become insufficient when task arrivals and resource states vary. Thus, Yang et al. \cite{yang2025task} introduce demand prediction before Multi Agent PPO (MAPPO)-based offloading in a satellite-HAP-terrestrial network, including terminal-to-HAP/LEO, inter-HAP, and HAP-to-LEO paths. Prediction allows the system to reserve resources and migrate tasks before congestion occurs. From the S$^2$C$^2$I perspective, the HAP is therefore not merely a relay between edge and cloud. It filters sensing data, stores delay-tolerant workloads, executes urgent inference, and forwards only overflow or globally valuable information to neighboring HAPs or satellites.

\subsubsection{Green, Distributed, and Emerging Stratospheric Computing}

The stratosphere also offers physical advantages for large-scale computing. Abderrahim et al. \cite{abderrahim2023leverage} propose data-center-enabled HAPs that exploit low ambient temperature for natural cooling and solar energy for server operation. Their analysis reports electricity-cost savings of about 12\% with one Data
Center-enabled HAP (DC-HAP) and 36\% with four HAPs in the studied configuration. They also identifying wireless-link capacity, server payload, workload heterogeneity, and maintenance as practical bottlenecks. Lu et al. \cite{lu2025distributed} extend this concept to multiple terrestrial data centers and DC-HAPs with heterogeneous tasks. Their results show that additional HAPs increase available computation and transmission opportunities, but also enlarge the optimization dimension. Thus, robust heuristics can be preferable to more accurate solvers in large topologies.
An emerging extension is quantum computing in the sky. Abderrahim et al. \cite{abderrahim2025green} place cryogenic processors on solar-powered HAPs and connect them through FSO and low-altitude repeaters. At the favorable operating altitude, the reduced thermal gradient lowers energy use by 21\% and supports 30\% more qubits than a terrestrial Quantum Data Center (QDC). Multi-HAP cooperation improves scalability, although radiation, quantum-link loss, payload reliability, and error correction remain open issues. Finally, the two-stage Multi-Agent DRL (MADRL) in \cite{liu2024distributed}, where HAP means hybrid access point. It is not a stratospheric architecture, but its decomposition of network-level energy control and device-level offloading is transferable to distributed HAP constellations. This distinction is important when importing algorithms across domains.

\begin{lessonslearned}

On-board computing should be designed as a constrained S$^2$C$^2$I service continuum. Task placement must be jointly determined with data locality, cache state, link availability, CPU/GPU capacity, queue stability, renewable-energy supply, and mission priority. Intelligence can predict workload and channel evolution, but its training and inference overhead must also be included in the payload budget. 
The central design objective is not simply to maximize executed cycles, but to convert sensed data into timely decisions with minimum communication, storage, computing, and energy cost.

\end{lessonslearned}

\subsection{Storage and Caching Hierarchy}

Storage and caching turn HAPs from transient forwarding nodes into persistent regional information hubs. In an S$^2$C$^2$I-integrated HAP system, storage does not only mean storing popular videos. It also includes high-definition maps, sensing observations, AI models, intermediate computing results, and delay-tolerant service packets. Therefore, storage directly couples with all other S$^2$C$^2$I functions: sensing generates data to be buffered and filtered; communication determines whether cached objects can be refreshed through feeder, inter-HAP, or HAP-UAV links; computing relies on locally available data libraries to avoid long-distance retrieval; and intelligence predicts content popularity, service demand, and cache replacement decisions. Representative HAP-oriented storage and caching designs are summarized in Table~\ref{tab:storage_caching_haps}.

\begin{table*}[!t]
\centering
\caption{Representative HAP Storage and Caching Designs and Their S$^2$C$^2$I Roles}
\label{tab:storage_caching_haps}
\renewcommand{\arrayrulewidth}{0.8pt}
\renewcommand{\tabcolsep}{3.0pt}
{\fontsize{7.1}{8.6}\selectfont
\begin{tabular}{m{0.75cm}<{\centering}||m{0.65cm}<{\centering}|m{2.3cm}<{\centering}|m{2.4cm}<{\centering}|m{3cm}<{\centering}|m{7.4cm}<{\centering}}
\hline \hline
\rowcolor{gray!15}
\textbf{Ref.} & \textbf{Year} & \textbf{Storage/Caching Paradigm} & \textbf{Link/Topology} & \textbf{S$^2$C$^2$I Role} & \textbf{Representative Insight}\\
\hline
\rowcolor{blue!10}
\cite{zhang2018air} & 2018 & Proactive content pushing and caching & HAPs-vehicles-RSUs & Storage-communication slicing & HAP broadcast rate, RSU unicast rate, and vehicle cache size exhibit service-dependent trading relationships\\
\hline
\rowcolor{blue!10}
\cite{vallero2022caching} & 2022 & HAPs MEC caching for urban & HAPs-ground users & Hot-content delivery and traffic offloading & HAPs caching is effective when simultaneously covered areas share similar content interests\\
\hline
\rowcolor{blue!10}
\cite{ren2022caching} & 2022 & HAPs data library& CAVs-RSUs-HAPs & Storage-computing-offloading coordination & RSU caches store fundamental data from the HAPs library, reducing HAPs-to-ground retrieval delay for local and RSU computing\\
\hline
\rowcolor{blue!10}
\cite{yang2023caching} & 2023 & Federated cache-update & CAVs-RSUs-HAPs-LEO & Distributed cache and computation control & Federated learning jointly updates caching, offloading, and computing-resource policies in satellite-HAP-terrestrial ITS\\
\hline
\rowcolor{green!10}
\cite{masood2021content} & 2021 & Hierarchical FL-based popularity prediction & Ground users-UAVs-HAP & Privacy-preserving cache intelligence & Users train locally, UAVs aggregate partial models, and the HAP performs global aggregation to predict contents suitable for UAV caching\\
\hline
\rowcolor{green!10}
\cite{yuan2023joint} & 2023 & Cache-enabled HAP/UAV & HAPs-UAVs-ground users & Hierarchical edge cache allocation & Joint cache placement, offloading, and HAP/UAV server selection reduce transmission delay under limited UAV storage and endurance\\
\hline
\rowcolor{green!10}
\cite{mu2024joint} & 2024 & Joint UAV deployment and cache control & HAPs-UAVs-users & Cache-aware aerial service placement & A hybrid-action DRL framework coordinates continuous UAV deployment and discrete caching decisions for low-latency content access\\
\hline
\rowcolor{orange!12}
\cite{kurt2021communication} & 2021 &HAPs storage for aerial delivery & HAPs-aerial delivery fleet-ground services & Communication-computing-caching-sensing support & HAPs storage helps maintain service data, delivery context, and local mission information for autonomous aerial delivery networks\\
\hline
\rowcolor{orange!12}
\cite{jaafar2022haps} & 2022 & HAPs-ITS storage unit for highways & HAPs-CAVs-highway sensors & Large regional service memory & HAPs storage is envisioned to support infotainment, fleet management, and temporary assistance when vehicle storage or processing fails\\
\hline

\rowcolor{purple!10}
\cite{alfattani2023multimode} & 2023 & Cache-aware HAPs payload & HAPs-SMBS/HAPs-RIS & Energy-aware storage-mode selection & Cached requests can activate the HAPs-SMBS mode, while cache misses can be served through relay/RIS modes\\
\hline
\rowcolor{purple!10}
\cite{zhang2026high} & 2026 & Multi-HAP caching and network& Data centers-HAPs-users & Cooperative cache and coded delivery & Dynamic HAPs cache placement, FSO backhaul routing, RF access beamforming, and multicast reduce long-term power cost\\

\hline \hline
\end{tabular}}
\vspace{2pt}
\end{table*}

\subsubsection{Proactive Content Pushing and User-Side Caching}

The first layer of the HAP storage hierarchy is formed by user-side and ground-edge caches, where HAPs exploit their wide-area broadcast capability to push contents before explicit requests arrive. In air-ground integrated vehicular networks, HAPs broadcast map or popular files to vehicles, while RSUs provide on-demand unicast service for cache misses \cite{zhang2018air}. 
For location-based map navigation, cache usefulness is determined by vehicle mobility and route evolution. For files of common interest, it depends on content popularity, generation, and expiration dynamics. Thus, storage becomes a slice-level resource rather than a passive buffer.
Urban HAPs caching extends this idea from vehicles to RAN traffic offloading. By equipping HAPs with a MEC server and activating multiple directional beams, cached popular contents can be delivered from the stratospheric layer to selected ground areas \cite{vallero2022caching}. The key insight is spatial demand correlation. If several beam-covered areas share similar popular contents, a single HAPs cache can offload more backhaul traffic; if the covered areas are highly heterogeneous, local ground MEC servers remain useful for area-specific popularity. Therefore, HAPs should cache regional hot objects, while RSUs, terrestrial MEC servers, vehicles, and user devices maintain more localized or short-lifetime contents.
From the S$^2$C$^2$I perspective, proactive pushing connects storage with communication and intelligence. Communication determines how much cached data can be broadcast or refreshed, while intelligence estimates mobility-aware and popularity-aware cache value. Sensing can further improve this process by providing road context, crowd density, emergency events, and traffic flow prediction. 

\subsubsection{Fundamental-Data Libraries for Computation-Aware Storage}

The second layer of storage is the HAP-side data library that supports edge computing. In intelligent transportation systems, many AI tasks require not only vehicle-generated input data but also fundamental data such as high-definition maps, weather information, and road conditions. A HAPs-assisted Intelligent Transportation Systems (ITS) framework therefore places a fundamental data library at the HAPs and allows RSUs to cache part of this library locally \cite{ren2022caching}. When a Connected and Autonomous Vehicle (CAV) selects local or RSU computing, cached fundamental data at the associated RSU avoids long-distance retrieval from the HAPs. When the requested data are not cached, the HAPs library provides a wide-area fallback. This architecture couples caching decisions with computation offloading, bandwidth allocation, and CPU allocation.
This coupling becomes more evident in satellite-HAP-terrestrial ITS systems. In \cite{yang2023caching}, LEO satellites, HAPs, RSUs, and CAVs jointly provide caching and computation offloading, where HAPs and RSUs act as distributed agents that update both execution and cache policies. The proposed federated cache-update matching mechanism indicates that caching cannot be optimized independently of computing placement. 
Thus, the storage hierarchy should be task-aware, not only popularity-aware.
HAPs-ITS studies for trans-continental highways further show that storage is a payload-level requirement. A HAP node may need tens of TBs of storage to support different CAV automation levels, fleet management, and temporary assistance when onboard vehicle storage or processing fails \cite{jaafar2022haps}. Similarly, HAPs-enabled aerial delivery networks require locally stored delivery context, navigation information, and service data to support low-latency autonomous fleet operation \cite{kurt2021communication}. These studies broaden the meaning of HAP storage from content caching to regional service memory. Within S$^2$C$^2$I, this memory enables computing to run close to data; reduces repeated feeder-link access; and provides resilience when terrestrial infrastructure is sparse or disrupted.

\subsubsection{Hierarchical HAP-UAV Caching and Learning}

The third layer is formed by collaborative storage between HAPs and UAVs. Since UAVs are close to users but constrained by energy, payload, and storage capacity, they are suitable for lightweight hot caches. HAPs, in contrast, offer larger coverage, more stable operation, and stronger aggregation capability. This naturally leads to a hierarchical cache architecture in which HAPs act as regional cache controllers or model aggregators, while UAVs serve as mobile edge cache points.
A representative learning-based design is the HAP-assisted multi-UAV caching framework using hierarchical Federated Learning (FL) \cite{masood2021content}. Users train local popularity-prediction models without uploading raw data, UAVs aggregate local updates, and the HAP performs global aggregation. This approach highlights the intelligence dimension of storage. The HAP does not merely store contents, but also stores and updates cache-prediction models. 
Resource allocation studies further examine how HAPs and UAVs should cooperate under limited cache capacity. HAPs provide stable wide-area service, while UAVs pre-store popular content closer to users \cite{yuan2023joint}. Joint optimization of cache placement, offloading decisions, and HAP/UAV server selection reduces transmission delay while respecting LAP endurance and storage constraints. In HAPs-assisted multi-UAV communications, the deployment position of UAVs and their cached files are jointly optimized through hybrid-action DRL \cite{mu2024joint}. This reveals another important S$^2$C$^2$I coupling. Storage placement changes the optimal aerial topology, while UAV deployment changes which cache is useful to which user. 

\subsubsection{Multi-HAP Cooperative Caching, Multicast, and Payload-Aware Storage}

Beyond a single HAP, storage can be extended into a cooperative aerial cache fabric. 
In multi-HAP rural connectivity, each caching-enabled HAP can use RF access links to serve ground users and FSO links for feeder and inter-HAP backhaul \cite{zhang2026high}. If a requested content is locally cached, the HAP serves it directly; otherwise, the request can be routed to another HAP or a data center through the FSO backhaul. 
In this architecture, storage, communication, and computing are optimized together. DRL determines dynamic cache placement, while convex optimization handles per-slot backhaul routing and RF access resource allocation.
Storage also influences HAPs' payload operating modes. A multimode HAP may switch among HAPs-Super 
Macro Base Station (SMBS), HAPs relay, and HAPs-RIS modes according to service demand, energy state, cache status, and computing availability \cite{alfattani2023multimode}. For a content request, if the object is already cached onboard, the HAPS-SMBS mode can directly deliver it; otherwise, relay or RIS modes may forward the content from the core network. If the pulled content is highly popular, the HAPs can proactively cache it for future requests. This shows that cache state can determine not only data routing but also the physical payload mode and energy consumption profile.
\begin{lessonslearned}
The storage and caching hierarchy in S$^2$C$^2$I-integrated HAP networks should be understood as a multi-layer data lifecycle. Ground devices and vehicles keep ultra-local temporary data; RSUs and UAVs cache near-user hot objects; HAPs maintain regional content libraries, model repositories, sensing buffers, and task-related fundamental data; and cloud or space-cloud layers preserve cold archives and global models. The main open challenge is to coordinate this hierarchy under limited payload, energy, backhaul, and freshness constraints. 

\end{lessonslearned}

\subsection{AI-based HAP Orchestration}
AI acts as the coordination plane that converts heterogeneous HAP observations into cross-layer actions. Rather than optimizing communication, computing, storage, sensing, or flight control independently, an S$^2$C$^2$I orchestrator must jointly interpret user demand, sensing products, task queues, cache states, platform motion, and renewable-energy availability. It then selects actions over multiple time scales, ranging from millisecond-level channel, beam, and power control to second-level association and offloading, and further to minute-or hour-level mission, cache, and model management. Representative AI-enabled HAP orchestration designs are summarized in Table~\ref{tab:ai_hap_orchestration}.

\begin{table*}[!t]
\centering
\caption{Representative AI-Based HAP Orchestration Designs and Their S$^2$C$^2$I Roles}
\label{tab:ai_hap_orchestration}

\renewcommand{\arrayrulewidth}{0.8pt}
\renewcommand{\tabcolsep}{3.0pt}

{\fontsize{7.1}{8.6}\selectfont
\begin{tabular}{
m{0.75cm}<{\centering}||
m{0.65cm}<{\centering}|
m{2.3cm}<{\centering}|
m{2.4cm}<{\centering}|
m{3cm}<{\centering}|
m{7.4cm}<{\centering}
}
\hline
\hline
\rowcolor{gray!15}
\textbf{Ref.}
& \textbf{Year}
& \textbf{AI/Control Paradigm}
& \textbf{Link/Topology}
& \textbf{S$^2$C$^2$I Role}
& \textbf{Representative Insight}\\
\hline

\rowcolor{blue!10}
\cite{dahrouj2023machine}
& 2023
& Unsupervised ensemble DNN
& Satellite-HAPs-ground
& User association and load balancing
& Ensemble DNNs approximate discrete scheduling decisions online without requiring labeled optimal solutions\\
\hline

\rowcolor{blue!10}
\cite{jain2025unleashing}
& 2025
& ANN with GAN/VAE data generation
& HAPs-IoT smart city
& Channel prediction and placement support
& Synthetic-data augmentation supports fast signal-strength prediction, with the Swish-based ANN reporting the lowest prediction error\\
\hline

\rowcolor{blue!10}
\cite{moon2025ai}
& 2025
& Online DNN and offline CNN
& Massive sensors-HAPs-ground
& Sensing reconstruction and model placement
& Online pointwise learning supports real-time adaptation, whereas offline end-to-end imaging improves reconstruction accuracy\\
\hline

\rowcolor{orange!12}
\cite{guan2019intelligent}
& 2019
& Q-learning with neural network
& HAPs massive-MIMO cells
& Autonomous channel allocation
& A neural approximator replaces the Q-table and enables adaptive channel allocation under high-dimensional network states\\
\hline

\rowcolor{orange!12}
\cite{anicho2021reinforcement}
& 2021
& RL versus swarm intelligence
& Multi-HAP coverage network
& Distributed platform coordination
& Swarm intelligence achieves more stable multi-HAP coverage than classical RL, which is sensitive to state-space partitioning\\
\hline

\rowcolor{orange!12}
\cite{abbasi2025ai}
& 2025
& K-means and DQN
& HAPs-multiuser
& Beam, frequency, and power control
& User clustering reduces scheduling complexity, while DQN improves power allocation over equal-power transmission\\
\hline

\rowcolor{orange!12}
\cite{sharma2025hap}
& 2025
& RL-based resource allocation
& HAPs-disaster users
& Adaptive bandwidth and coverage control
& The RL controller adapts bandwidth and coverage according to changing user demand and link conditions in disaster areas\\
\hline

\rowcolor{orange!12}
\cite{ince2026aoi}
& 2026
& Modified MADDPG
& HAPs-aided vehicular networks
& AoI, reliability, and energy-aware control
& Local and global critics coordinate platoon leaders to reduce AoI and power consumption while improving delivery reliability\\
\hline

\rowcolor{orange!12}
\cite{li2026efficient}
& 2026
& EPG-enhanced MAPPO
& LEO-HAPs-maritime users
& Multi-tier offloading and resource allocation
& EPG rewards align decentralized agents with system-wide latency, energy, and cost objectives\\
\hline

\rowcolor{green!10}
\cite{kiam2020ai}
& 2020
& PDDL planning with ALNS
& Multi-HAP sensing missions
& Weather-aware task and motion planning
& Symbolic-numeric planning jointly captures platform dynamics, time-varying weather, and heterogeneous sensing tasks\\
\hline

\rowcolor{green!10}
\cite{han2025agent}
& 2025
& LLM-based multi-agent system
& Multi-HAP coverage network
& Event-aware autonomous coordination
& In-context learning, fine-tuning, retrieval, and tool calling enable HAP agents to interpret events and coordinate coverage\\
\hline

\rowcolor{green!10}
\cite{xing2026generative}
& 2026
& Generative-AI agent
& HAP propulsion and communication systems
& Cross-domain energy and beam control
& Agent-assisted modeling exposes the coupling among propulsion power, communication QoS, and energy-efficient beamforming\\
\hline

\rowcolor{green!10}
\cite{yan2026hierarchical}
& 2026
& Hierarchical LLM-RL control
& HAPs-UAV-terrestrial network
& Mobility, handover, and connectivity control
& A HAP-side LLM performs slow-timescale planning, while local LLM-RL agents execute fast motion and association decisions\\
\hline
\hline
\end{tabular}}

\vspace{2pt}
\end{table*}

\subsubsection{Data-Driven State Prediction and Fast Network Control}

Predictive models reduce the need to repeatedly solve complex HAP optimization problems from scratch. In integrated satellite-HAP-ground networks, Dahrouj et al.  \cite{dahrouj2023machine} use an unsupervised ensemble of Deep Neural Networks (DNNs) to determine whether each user should associate with a terrestrial base station or a HAP. The objective and constraints are embedded into the training loss, allowing online scheduling without labeled optimal solutions. Jain et al.  \cite{jain2025unleashing} instead learn the radio environment itself. An Artificial Neural Network (ANN) predicts HAP-to-IoT signal strength from altitude, distance, frequency, transmit power, and propagation features, while Generative Adversarial Networks (GANs) and Variational Auto-Encoders (VAEs) enlarge the available dataset. Such predictors can support proactive placement, beam selection, handover, and coverage repair before the link deteriorates.
Learning can also transform raw sensing observations into mission-ready information. The framework in \cite{moon2025ai} reconstructs spatially correlated measurements from massive simultaneous sensor transmissions. Its model-driven pointwise estimator is trained online for real-time adaptation, whereas its end-to-end Convolutional Neural Network (CNN) is trained offline to obtain higher imaging accuracy. Terrestrial BSs collect and preprocess training data and then transfer models or samples to the HAP. This division illustrates an S$^2$C$^2$I model lifecycle. Sensing produces correlated measurements; communication transports aggregated signals and updates; computing performs reconstruction; storage retains training data and models; and intelligence chooses the appropriate online/offline inference mode.
At the physical layer, prediction must be combined with structured control. Abbasi et al.~\cite{abbasi2025ai} first cluster users by the K-means algorithm \cite{luo2025weighted}, allocate orthogonal subcarriers within each cluster, and then use Deep Q-Network (DQN) for HAP power allocation. The hybrid design is preferable to an unconstrained end-to-end learner because clustering and codebook beamforming preserve domain structure, while learning handles the dynamic component. 

\subsubsection{Reinforcement Learning and Multi-Agent Cross-Layer Orchestration}

Reinforcement Learning (RL) is attractive when HAP decisions affect future channels, queues, energy, and mobility. Early work combines Q-learning with a back-propagation neural network for HAP massive-MIMO channel allocation \cite{guan2019intelligent}. However, the comparison in \cite{anicho2021reinforcement} shows that classical RL can suffer from state discretization, coverage dips, and slow convergence. A swarm-intelligence baseline provides substantially more stable multi-HAP coverage. Therefore, learning should not automatically replace low-complexity heuristics, especially when a HAP must satisfy hard availability or safety constraints.
Recent MADRL designs address distributed and partially observable systems. In HAP-aided vehicular networks, each platoon leader acts as an agent, while local and global critics jointly optimize Age of Information (AoI), delivery probability, and power consumption~\cite{ince2026aoi}. In maritime satellite-HAP networks, MAPPO combines centralized training and decentralized execution with Exact Potential Game (EPG) rewards so that individual offloading decisions improve a common latency-energy-cost objective \cite{li2026efficient}. For disaster recovery, PPO adjusts bandwidth and coverage according to changing user density and link conditions \cite{sharma2025hap}. Interference management in HAP-enabled vertical heterogeneous networks similarly motivates distributed RL because centralized optimization requires rapidly changing global channel information \cite{shamsabadi2025interference}.
These works indicate that reward design is the key S$^2$C$^2$I interface. A reward based only on throughput may overload the HAP processor, drain propulsion reserves, discard valuable sensed data, or create stale decisions. Instead, it should include communication rate and reliability, sensing utility or AoI, task latency, cache-hit value, CPU/GPU occupancy, and propulsion/communication energy. Additionally, model-based schemes remain useful as teachers, benchmarks, and safety fallbacks. For example, the local-RSU-HAP framework in \cite{ren2022caching} jointly optimizes caching, offloading, bandwidth, and computing resources. The aerial-marine search-and-rescue design in \cite{wang2026computation} decomposes server selection, transmit power, and CPU allocation through matching and convex optimization. Such solutions provide feasible action masks, expert demonstrations, or recovery policies when a learned controller becomes uncertain.

\subsubsection{Planning, LLM Agents, and Hierarchical Autonomy}

Long-horizon HAP operation requires reasoning over weather, mission semantics, and heterogeneous objectives that are difficult to encode as a single Markov reward. Kiam et al.  \cite{kiam2020ai} formulate multi-HAP mission planning in Problem Domain Definition Language (PDDL) and combine a domain-independent planner with Adaptive Large Neighborhood Search (ALNS). Unlike a separated task-then-motion pipeline, their common search space accounts for continuous HAP dynamics, time-varying wind, exogenous events, and sensor-dependent monitoring tasks. This model-based planning layer is suitable for mission feasibility and constraint enforcement, but its computation grows with the number of HAPs, tasks, and environmental events.
Large Language Model (LLM) agents extend orchestration from numeric state optimization to semantic event interpretation \cite{luo2025toward}, \cite{sun2025edge}, \cite{luo2026location}. This framework equips multiple HAP agents with in-context learning, fine-tuning, information retrieval, and tool calling \cite{han2025agent}. It can coordinate coverage and react to natural-language descriptions of unexpected events. A more hierarchical design is developed in \cite{yan2026hierarchical}. A HAP-side LLM performs slower load-balancing and handover planning, while each UAV combines a slow LLM spatial reasoner with a fast RL controller for motion and connectivity. This separation matches the natural time scales of S$^2$C$^2$I operation and avoids invoking a large model for every low-level action.
Generative AI can further assist cross-domain modeling. Xing et al. \cite{xing2026generative} use an interactive agent to derive a propulsion model that includes hull-propeller interference and then formulate energy-efficient communication beamforming. Importantly, the generative model mainly supports multidisciplinary reasoning and problem construction. This distinction suggests a practical architecture: LLMs interpret missions, retrieve models, decompose objectives, and select tools; lightweight predictors, optimization solvers, and DRL policies execute bounded actions; and a rule-based safety layer verifies energy, flight, and spectrum constraints before actuation.

\begin{lessonslearned}
AI-based HAP orchestration is most effective as a hierarchical hybrid system rather than a single end-to-end model. Predictive learning compresses complex states; optimization and fast RL execute resource decisions; MARL coordinates distributed HAP, UAV, satellite, vehicle, and gateway agents; and planners or LLMs handle long-horizon missions and semantic events. The optimization objective must explicitly couple all S$^2$C$^2$I functions and include the communication, computation, storage, and energy overhead of AI itself. 
\end{lessonslearned}

\section{Standardization, Open Resources, and Evaluation}\label{Sec-V} 
The transition of HAPs from communication relays to S$^2$C$^2$I-integrated infrastructures depends not only on algorithmic advances, but also on spectrum authorization, interoperable network interfaces, aviation certification, and reproducible software. And it is also very important to have an evaluation method that captures the coupling among sensing, storage, communication, computing, and intelligence. Existing efforts are distributed across different communities. ITU-R primarily regulates spectrum sharing and interference; 3GPP specifies cellular NTN access and core-network integration; ETSI addresses service deployment; IEEE and O-RAN provide aerial-network and programmable-RAN building blocks; and aviation-oriented industry groups address certification, traffic management, and operational risk. Consequently, no single framework currently standardizes the complete data-decision-actuation lifecycle of an S$^2$C$^2$I-enabled HAP. This section therefore synthesizes these complementary activities, surveys reusable research resources and real trials, and proposes a common evaluation methodology.

\subsection{Standardization and Regulatory Landscape}
Table~\ref{tab:haps_standardization} summarizes the current landscape, which is elaborated as follows:

\begin{table*}[!t]
\centering
\caption{Representative Standardization and Regulatory Activities Relevant to S$^2$C$^2$I-Integrated HAPs}
\label{tab:haps_standardization}
\renewcommand{\arrayrulewidth}{0.8pt}
\renewcommand{\tabcolsep}{3.4pt}
{\fontsize{7.2}{8.7}\selectfont
\begin{tabular}{m{1.55cm}<{\centering}|m{3.15cm}<{\centering}|m{4.15cm}<{\centering}|m{3.25cm}<{\centering}|m{4.05cm}<{\centering}}
\hline\hline
\rowcolor{gray!15}
\textbf{Body/Activity} & \textbf{Representative Document or Release} & \textbf{Main Scope} & \textbf{S$^2$C$^2$I Relevance} & \textbf{Remaining HAP-Specific Gap}\\
\hline
ITU/WRC & WRC-19 HAPs \cite{itu2025haps}; WRC-23 HIBS \cite{itu2023wrc23} & Spectrum identification, coexistence, link directions, power-flux-density and interference-protection conditions & Spectrum for service, and links; indirect constraints on sensing and data transport & Authorization, cross-border coordination, dynamic sharing are fragmented\\
\hline
3GPP & TR~38.811 \cite{3gpp38811}, TR~38.821 \cite{3gpp38821}, Releases 17--19 NTN \cite{3gpp2025ntn} & NTN access, mobility, backhaul, onboard UPF, local switching, and store-and-forward & Communication; partial storage and computing support & Quasi-stationary HAP profiles, HAP-UAV interfaces, platform-state exposure, and sensing metadata\\
\hline
ETSI & GS MEC~012 \cite{etsiMEC012}, GR MEC~047 \cite{etsiMEC047}, GR NFV-EVE~023 \cite{etsiNFVEVE023} & Edge APIs, distributed MEC, HAP as NFV, lightweight virtualization, and failure management & Computing, storage, communication, and orchestration & Models for HAP availability, accelerator capability, energy state, and lifecycle operations\\
\hline
O-RAN & O-RAN architecture, near-RT and non-RT RIC reference software \cite{oranArchitecture,oranNonRTRIC} & Open RAN interfaces, telemetry, policies, xApps/rApps, AI/ML training, deployment, inference, and feedback & Intelligence-driven communication control and a reusable hierarchy & Models for sensing tasks, cache/model freshness, flight dynamics, energy state, or HAP-to-HAP coordination\\
\hline
IEEE & IEEE~1920.1-2022 \cite{ieee1920_1} & Self-organized aerial-network architecture, security, and data models & Inter-HAP and HAP-UAV communication foundations & No integrated cloud, storage, sensing, or mission-autonomy profile\\
\hline
HAPS Alliance& Certification, risk, payload, traffic-management, and flight-test white papers \cite{hapsCertification,hapsRisk,hapsPayload,hapsFleet,hapsFlightTests} & Airworthiness, operational risk, environmental qualification, and autonomous fleet operation& Cross-cutting constraints on every S$^2$C$^2$I function and on platform endurance & Industry guidance rather than binding global standards\\
\hline\hline
\end{tabular}}
\end{table*}

\subsubsection{Spectrum Regulation and HIBS Operation}

The ITU provides the global regulatory foundation for HAP radio systems. For HAPs, the World Radiocommunication Conference 2019 (WRC-19) identified 31-31.3 GHz and 38-39.5 GHz for worldwide use, retained the worldwide identifications at 47.2-47.5 GHz and 47.9-48.2 GHz \cite{itu2025haps}. These allocations are especially relevant to high-capacity feeder and inter-HAP links. However, their practical use remains constrained by rain attenuation, antenna pointing, power-flux-density limits, and coexistence with incumbent fixed, mobile, and satellite services.
A complementary regulatory concept is the HAP as an International Mobile Telecommunications (IMT) Base Station (HIBS) \cite{euler2021primer}, under which a HAP reuses terrestrial IMT bands and can directly serve conventional mobile terminals. WRC-23 expanded the regulatory basis for HIBS operation in the 2 GHz and 2.6 GHz ranges and established associated protection conditions \cite{itu2023wrc23}. This development is important for direct-to-device HAP access because it reduces the need for specialized terminals and allows spectrum and device ecosystems to be shared with terrestrial networks. Nevertheless, an international identification does not by itself authorize deployment. National licensing, cross-border coordination, emission masks, aviation approvals, and protection of incumbent services remain necessary. Therefore, HAP evaluations should explicitly distinguish between \emph{technically feasible spectrum}, \emph{internationally identified spectrum}, and \emph{nationally licensed spectrum}.

\subsubsection{3GPP NTN Evolution}

3GPP provides the main cellular standardization path for integrating airborne and spaceborne nodes into 5G systems. TR~38.811 \cite{3gpp38811} established NTN deployment scenarios, channel assumptions, timing relationships, and transparent and regenerative payload models, while TR~38.821 studied corresponding NR solutions \cite{3gpp38821}. Release 17 introduced the first normative NR-NTN and IoT-NTN functions. Release~18 subsequently enhanced uplink coverage, network-assisted verification of terminal location, NTN-Gornud and NTN-NTN mobility, operation above 10~GHz, satellite backhaul, and architectural support for onboard UPF and local data switching \cite{3gpp2025ntn}. Release 19 further considers regenerative payloads, RedCap support, store-and-forward operation, and direct user-to-user communication through the non-terrestrial segment.
These mechanisms are directly reusable by HAP systems. Transparent payload specifications match bent-pipe HAP relays, whereas regenerative payloads support onboard gNB, routing, edge processing, and local traffic breakout. Mobility signaling and conditional handover are useful when a platform drifts, changes its footprint, or transfers service among multiple HAPs. Store-and-forward functions can also support delayed sensing uploads and intermittent feeder links. However, most normative NTN profiles and reference parameters have been developed around satellite systems. HAP-specific profiles for quasi-stationary drift, stratospheric channels, HAP-UAV links, platform energy states, and multi-functional payload orchestration remain incomplete. In particular, 3GPP does not yet define common interfaces for exposing sensing quality, cache/model freshness, onboard accelerator availability, or flight-energy reserves to the network scheduler.

\subsubsection{MEC, NFV, and Programmable RAN}

ETSI MEC and Network Functions Virtualization (NFV) offer reusable abstractions for deploying communication and computing functions \cite{sun2019low, yu2011cost, sun2020dynamic}. MEC radio-network information services expose radio measurements to edge applications, while distributed MEC architectures support application placement across multiple edge sites \cite{etsiMEC012,etsiMEC047}. More directly, ETSI GR NFV-EVE~023 treats HAPs as resource- and power-constrained NFV infrastructure and studies scaling network services onto HAPs resources \cite{etsiNFVEVE023}. 
This is an important step from HAP as a radio relay toward HAP as a mobile mini-data center.
O-RAN provides another useful, although not HAP-specific, control framework. Its near-real-time RAN Intelligent Controller (RIC) supports fine-grained control through xApps, whereas the non-real-time RIC supports policy optimization, analytics, AI/ML model training and distribution \cite{oranArchitecture,oranNonRTRIC}. These timescales map naturally to the hierarchical intelligence described in Section~\ref{Sec-IV}: fast onboard control can reside in a near-real-time RIC or local controller, while slower cross-HAP policy and model management can be handled by the non-real-time RIC. Nevertheless, O-RAN information models currently focus on RAN entities and do not natively describe flight dynamics, atmospheric observations, cache state, sensing tasks, or propulsion constraints. A HAP-oriented service-management layer must therefore augment standard RAN telemetry with platform and S$^2$C$^2$I context.

\subsubsection{Aerial Networking, Certification, and Safety}

IEEE 1920.1-2022 specified architecture, security, and data models for self-organized air-to-air communication among aerial nodes \cite{ieee1920_1}. It is relevant to inter-HAP and HAP-UAV networking, although it was a trial-use standard and became inactive-reserved in 2026. Its status illustrates a broader issue: aerial-network standards have not yet converged into an active end-to-end framework that jointly covers communication, onboard cloud functions, and mission autonomy.
Telecommunication interoperability alone is insufficient for operational deployment. The HAPs Alliance has published guidance on certification pathways, acceptable risk levels, payload operation in the stratosphere, collaborative traffic management, and flight communication tests \cite{hapsCertification}, \cite{hapsRisk}, \cite{hapsPayload}, \cite{hapsFleet}, \cite{hapsFlightTests}. These documents are not formal international standards, but they provide engineering evidence and aviation-oriented terminology that are largely absent from communication specifications. For example, certification must consider the complete aircraft--payload system, command-and-control continuity, detect-and-avoid functions, failure containment, environmental qualification, and risk imposed on third parties. Hence, S$^2$C$^2$I orchestration should be treated as a safety-relevant system. A learning policy that improves throughput but violates thermal, energy, geofencing, or flight-control margins is not deployable.


\begin{lessonslearned}
Current standardization is sufficiently mature to assemble a communication-centric HAP prototype, but not to guarantee interoperability of a fully integrated S$^2$C$^2$I platform. The most urgent missing elements are a common HAP capability descriptor, telemetry models for energy/thermal/compute/storage/sensing state, northbound APIs for mission intent, and cross-domain assurance procedures connecting telecom orchestration to flight safety. Future standards should avoid defining five independent function silos; instead, they should specify how resource and service states are exchanged across them.
\end{lessonslearned}

\subsection{Open Software, Data, Testbeds, and Field Evidence}

Reproducible HAP research is difficult because no open platform currently models the aircraft, atmosphere, radio access, feeder/backhaul network, onboard computing/storage, AI control, and mission-level safety in one validated stack. Nevertheless, a useful experimental environment can be assembled by combining resources from terrestrial 5G, satellite networking, radio-channel modeling, cloud orchestration, weather reanalysis, and aerial testbeds. Table~\ref{tab:haps_open_resources} classifies representative resources by their role and HAP readiness.

\subsubsection{Communication and Channel Modeling}
5G-LENA \cite{fiveGLena} is an open-source, 3GPP-aligned NR module for NS-3 that models the full protocol stack and supports end-to-end system evaluation. It is suitable for HAP service-link scheduling, interference, handover, traffic, and multi-cell studies. However, HAP altitude, stratospheric propagation, feeder links, and platform motion must be added explicitly. OpenAirInterface (OAI) \cite{oaiNTN} complements simulation with executable users, BSs, and core-network software. Its NTN implementation supports Release~17 adaptations for GEO/LEO scenarios and therefore provides a strong basis for HAP protocol emulation after timing, Doppler, ephemeris, and channel parameters are adapted.
At the link and propagation levels, Sionna \cite{sionna} offers differentiable, GPU-accelerated physical layer and ray-tracing components. This makes it useful for AI-native waveform, beamforming, localization, and digital-twin studies. QuaDRiGa  \cite{quadriga} generates realistic MIMO channel impulse responses and includes satellite-oriented configurations and an open non-GEO satellite extension \cite{quadrigaSatellite}. Neither tool by itself captures stratospheric weather, aircraft vibration, platform attitude errors, or complete RAN behavior. Therefore, their outputs should be calibrated with measurements or combined with analytical HAP channel models rather than treated as direct flight-test substitutes.

\subsubsection{Transport, Cloud, and Orchestration}

OpenSAND \cite{opensand} emulates satellite systems and can interconnect real applications and equipment, making it useful for feeder-link impairments, gateway switching, and hybrid space-HAP transport experiments. Hypatia  \cite{hypatia} precomputes time-varying satellite topologies and performs packet-level NS-3 simulations. Although neither is HAP-native, they can represent the space segment above the HAP layer and study cross-layer routing or cloud access.
For onboard and distributed service orchestration, ETSI Open Source Management and Orchestration (MANO) (OSM) \cite{etsiOSM} provides an open NFV-MANO implementation aligned with ETSI NFV. O-RAN components provide near-RT/non-RT RIC reference implementations \cite{oranSoftware}. These platforms can instantiate communication, inference, sensing-preprocessing, and caching functions as containers. Their main limitation is semantic rather than computational. Researchers should define custom descriptors and telemetry exporters for HAP location, battery state, accelerator occupancy, cache contents, and feeder availability.

\subsubsection{Atmospheric and Energy Traces}

Platform endurance and link availability should be evaluated under time-correlated environmental traces. ERA5 \cite{era5} provides global hourly reanalysis data, including pressure-level wind and atmospheric variables, and is suitable for trajectory, station-keeping, temperature, and cloud-related scenario generation. NASA POWER  \cite{nasaPower} provides globally accessible solar and meteorological parameters through web services and APIs. These datasets enable trace-driven energy harvesting and propulsion studies, but their spatial and temporal resolutions may be too coarse for local turbulence or cloud blockage. A defensible evaluation should report the dataset version, variables, pressure level, interpolation method, location, season, and any bias correction against local observations.

\subsubsection{Testbeds and Field Trials}

AERPAW  \cite{aerpaw} provides a digital twin, aerial nodes, outdoor UAV experimentation, and public datasets. Colosseum \cite{colosseum} provides 256 programmable software radios for large-scale, repeatable wireless emulation. They cannot reproduce stratospheric aerodynamics or long-duration solar operation, but they are valuable intermediate steps for validating sensing, localization, O-RAN applications, and task offloading before expensive HAP flights.
Additionally, public stratospheric trials provide essential reality checks. In 2020, HAPSMobile and Loon demonstrated LTE from the fixed-wing Sunglider  \cite{softbank2024hawk30} above 62,000 ft. The payload maintained LTE connectivity for approximately 15 hours, used a 5-MHz LTE Band 28 service link and an mmWave feeder link. It can support a video call using conventional smartphones under temperatures down to $-73^{\circ}$C and winds exceeding 58 knots. In 2025, AALTO's Zephyr \cite{aalto2025zephyr} flew continuously above 60,000 ft over Kenya for several days and connected a ground 4G device and gateway, including a telephone call routed through the platform. These trials establish the feasibility of stratospheric direct-to-device access. However, public releases generally provide only selected headline results rather than raw channel traces, energy telemetry, full antenna patterns, or reproducible configurations. Thus, they should be used to bound assumptions and validate orders of magnitude, not as complete public benchmarks.

\begin{table*}[!t]
\centering
\caption{Representative Open Resources for HAP Research}
\label{tab:haps_open_resources}
\renewcommand{\arrayrulewidth}{0.8pt}
\renewcommand{\tabcolsep}{3.2pt}
{\fontsize{7.15}{8.6}\selectfont
\begin{tabular}{m{2.15cm}<{\centering}|m{2.05cm}<{\centering}|m{2.0cm}<{\centering}|m{5.0cm}<{\centering}|m{5.25cm}<{\centering}}
\hline\hline
\rowcolor{gray!15}
\textbf{Resource} & \textbf{Primary Layer} & \textbf{HAP Readiness} & \textbf{Recommended Use} & \textbf{Main Limitation for S$^2$C$^2$I Evaluation}\\
\hline
5G-LENA \cite{fiveGLena} & Full-stack NR simulation & Adaptable & Service-link scheduling, multi-cell interference, mobility, QoS, traffic, and repeatable Monte Carlo evaluation & No native stratospheric platform, feeder-link, flight-energy, sensing, storage, or HAP-specific channel model\\
\hline
OpenAirInterface NTN \cite{oaiNTN} & Executable RAN/core & NTN-ready & End-to-end user-BS-core prototyping, protocol conformance experiments, channel-emulator & Existing NTN profiles are primarily satellite-oriented; HAP timing, drift, channel, and platform-state interfaces require extension\\
\hline
Sionna \cite{sionna} and QuaDRiGa \cite{quadriga,quadrigaSatellite} & Physical, channel, and digital twin & Adaptable & Beamforming, MIMO/FSO abstractions, differentiable optimization, and geometry/channel generation & Incomplete atmosphere, vibration, pointing, aircraft attitude, network queues, and application/service modeling\\
\hline
OpenSAND \cite{opensand}, Hypatia \cite{hypatia} & Transport/network emulation & Adaptable & Satellite-HAP-gateway transport, feeder impairments, gateway switching, time-varying space topology, routing, and real-application tests & Designed for satellite systems; no onboard HAP payload, propulsion, solar-energy, sensing, or cache/compute model\\
\hline
O-RAN \cite{oranSoftware} and ETSI OSM \cite{etsiOSM} & RAN/NFV orchestration & Building block & xApp/rApp development, policy control, model distribution, container lifecycle, telemetry, auto-scaling, and service migration & HAP resource descriptors, mission intent, safety constraints, and S$^2$C$^2$I telemetry schemas are absent\\
\hline
ERA5 \cite{era5} and NASA POWER \cite{nasaPower} & Environmental data & Input traces & Wind, temperature, solar radiation, seasonal scenarios, station-keeping load, harvested-energy, and link-availability traces & Gridded reanalysis/model data may miss local turbulence and clouds; pressure-level selection, interpolation, and validation must be reported\\
\hline
AERPAW \cite{aerpaw} and Colosseum \cite{colosseum} & Testbed/emulation & Proxy validation & UAV links, sensing/localization, digital-twin-to-field transfer, O-RAN control, and repeatable large-scale radio experiments & Low-altitude or indoor emulation cannot reproduce stratospheric aerodynamics, endurance, temperature, pressure, or long-range propagation\\
\hline
Zephyr \cite{aalto2025zephyr} and  Sunglider \cite{softbank2024hawk30} & Stratospheric field evidence & Direct evidence & Validate feasible altitude, commercial-device access, service/feeder architectures, environmental ranges & Raw traces, configurations, energy budgets, complete KPIs, and negative/failure results are generally not publicly released\\
\hline\hline
\end{tabular}}
\vspace{2pt}
\end{table*}
\begin{lessonslearned}
A practical open evaluation chain is therefore modular: ERA5 and NASA POWER generate environmental and energy traces; QuaDRiGa or Sionna generate channel realizations; 5G-LENA evaluates network-scale behavior; OAI and channel emulators validate protocol implementations; OpenSAND/Hypatia model the upper transport segment; O-RAN SC/OSM implement policy and lifecycle control; and AERPAW or Colosseum provides hardware validation. The interfaces among these tools should exchange timestamped platform state, channel state, queues, resource availability, task metadata, and control actions. Without a consistent time base and state schema, independently validated modules may still produce an invalid end-to-end result.
\end{lessonslearned}

\subsection{Evaluation Framework for S$^2$C$^2$I-Integrated HAPs}
 \begin{figure}[!t]
\centering
 \includegraphics[width=3.5 in]{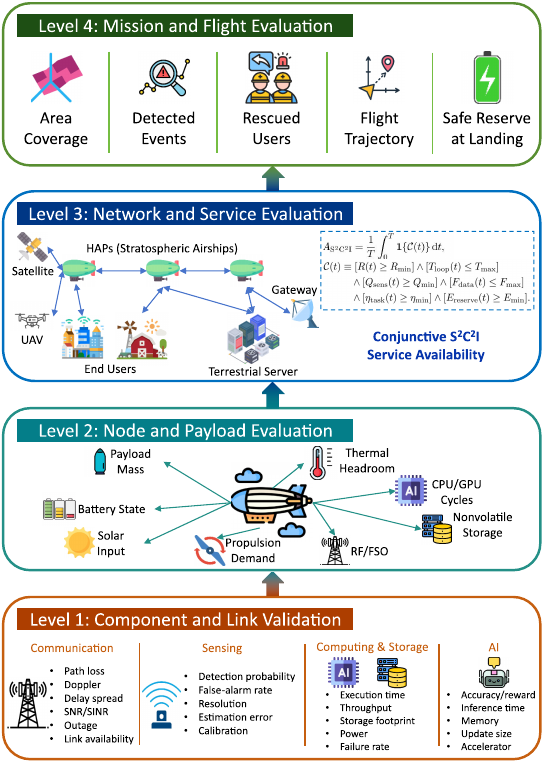}
   \caption{Four-level evaluation framework for S$^2$C$^2$I-integrated HAPs, spanning component validation, node-level resource feasibility, network service performance, and mission-level effectiveness.}
\label{fig:evaluation}
   \vspace{-0.5cm}
\end{figure}

Traditional HAP studies often optimize one metric, such as coverage \cite{lou2025coverage}, rate \cite{wang2021delay}, or energy \cite{liu2025energy}, while holding other functions fixed. Such an evaluation is insufficient for S$^2$C$^2$I systems because improvements can be displaced across domains. For example, onboard compression reduces feeder traffic but consumes compute energy and may reduce sensing accuracy; caching improves latency but occupies storage and requires refresh traffic; and a larger AI model may improve decision quality while increasing inference delay, thermal load, and model distribution overhead. We therefore propose a four-level evaluation hierarchy, as shown in Fig. \ref{fig:evaluation}.

\subsubsection{Level 1: Component and Link Validation}

The first level validates individual physical and software components. Communication results should report path loss, Doppler, delay spread, SNR/SINR, outage, and link availability across altitude, elevation angle, weather, and platform attitude. Sensing results should report detection probability, false-alarm rate, resolution, estimation error, and calibration. Computing and storage modules should be profiled using execution time, throughput, memory/storage footprint, power, thermal behavior, and failure rate on the actual or representative onboard hardware. AI modules should report accuracy or reward together with inference time, memory, accelerator utilization, update size, and robustness to distribution shift. This level prevents unrealistically abstract component models from dominating network conclusions.

\subsubsection{Level 2: Node and Payload Evaluation}

The second level evaluates resource contention within one HAP. Experiments should jointly account for payload mass, average and peak electrical power, battery state of charge, solar input, propulsion demand, thermal headroom, CPU/GPU cycles, nonvolatile storage, and RF/optical front-end use. The principal question is not whether each module works independently, but whether the entire payload remains feasible under concurrent sensing, communication, computing, caching, and inference. We should report resource-utilization time series and the frequency and duration of constraint violations, rather than only average utilization.

\subsubsection{Level 3: Network and Service Evaluation}

The third level considers multiple HAPs, satellites, UAVs, gateways, terrestrial MEC sites, and users. Besides throughput and delay, evaluation should include coverage continuity, handover interruption, feeder-link utilization, queue stability, task-completion ratio, cache hit ratio, model/cache freshness, service migration overhead, control-plane traffic, and failure recovery. 
Multi-HAP studies should also vary gateway density, HAP spacing, weather correlation, inter-HAP capacity, and common-cause failures.
To avoid declaring success when only one subsystem meets its target, we define a conjunctive S$^2$C$^2$I service availability:
\begin{equation}
\begin{aligned}
A_{\mathrm{S^2C^2I}}&=\frac{1}{T}\int_{0}^{T}\mathbf{1}\!\left\{\mathcal{C}(t)\right\}\mathrm{d}t,\\
\mathcal{C}(t)&\equiv [R(t)\geq R_{\min}]\wedge[T_{\mathrm{loop}}(t)\leq T_{\max}]\\
&\quad\wedge[Q_{\mathrm{sens}}(t)\geq Q_{\min}]\wedge[F_{\mathrm{data}}(t)\leq F_{\max}]\\
&\quad\wedge[\eta_{\mathrm{task}}(t)\geq \eta_{\min}]
\wedge[E_{\mathrm{reserve}}(t)\geq E_{\min}].
\end{aligned}
\label{eq:s2c2i_availability}
\end{equation}
where \(T\) is the mission-evaluation horizon, and \(A_{\mathrm{S^2C^2I}}\) denotes the fraction of time during which the integrated HAP service remains valid. 
The indicator \(\mathbf{1}\{\mathcal{C}(t)\}\) equals one only when the joint condition \(\mathcal{C}(t)\) is satisfied, and equals zero otherwise. 
Here, \(R(t)\) and \(R_{\min}\) denote the instantaneous effective communication rate and its minimum requirement; \(T_{\mathrm{loop}}(t)\) and \(T_{\max}\) denote the closed-loop sensing-storage-communication-computing-intelligence latency and its maximum tolerable bound; \(Q_{\mathrm{sens}}(t)\) and \(Q_{\min}\) denote the sensing quality and its minimum acceptable threshold; \(F_{\mathrm{data}}(t)\) and \(F_{\max}\) denote the staleness of data, cached content, or AI models and its maximum tolerable bound;  $\eta_{\mathrm{task}}(t)$ and $\eta_{\min}$
denote the task-completion ratio and its minimum required level; and \(E_{\mathrm{reserve}}(t)\) and \(E_{\min}\) denote the protected onboard energy reserve and the minimum safety threshold. 
The logical operator \(\wedge\) indicates that all constraints must be satisfied simultaneously. 
Thus, this metric evaluates conjunctive S$^2$C$^2$I service availability rather than a weighted average performance score, preventing the success of one function from masking the failure of another.


\subsubsection{Level 4: Mission and Flight Evaluation}

The fourth level evaluates whether network improvements translate into mission value without compromising the aircraft. Representative metrics include completed observations, detected events, rescued/served users, area-time coverage, delivered useful information, mission duration, safe reserve at landing, probability of loss of service, and recovery time after failures. Communication energy must be reported together with propulsion and avionics energy because the latter often dominates the platform budget. Similarly, an algorithm that extends network lifetime but forces large trajectory deviations or decreases safety margin should not be considered energy-efficient at the mission level.

\subsection{Case Study: S$^2$C$^2$I-Enabled HAP-SAGIN for Emergency Response}

 \begin{figure*}[!t]
\centering
 \includegraphics[width=7 in]{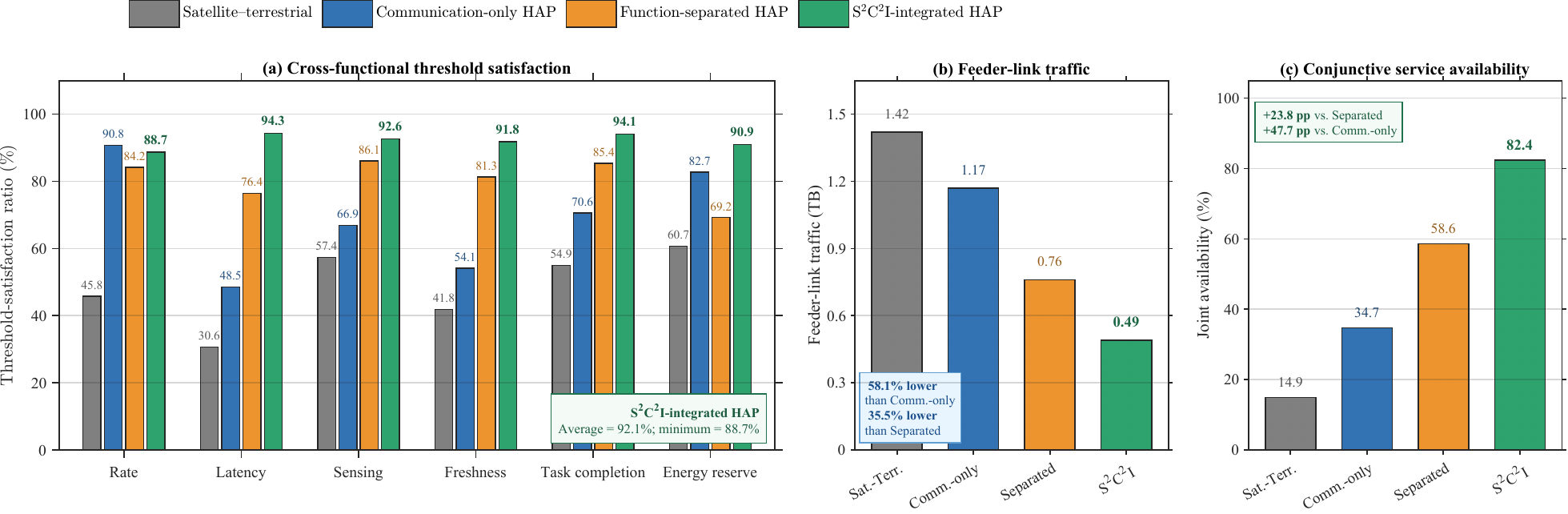}
   \caption{Case-study performance comparison of different
SAGIN architectures. (a) Satisfaction
ratios of cross-functional requirements. (b) Aggregate
feeder-link traffic. (c) Conjunctive S$^2$C$^2$I service
availability. The integrated HAP architecture achieves a more
balanced performance profile, substantially reduces feeder-link traffic, and improves the probability of simultaneously satisfying all service requirements.}
\label{fig:case}
   \vspace{-0.5cm}
\end{figure*}

To demonstrate the system-level gain of S$^2$C$^2$I integration, we consider an emergency-response HAP-SAGIN benchmark where terrestrial BSs and backhaul links are partially damaged after a wildfire or earthquake. The benchmark is implemented in NS-3 and evaluated on a PC equipped with an Intel Core i9-14900HX CPU at 2.2~GHz and an NVIDIA RTX 4070 GPU with 8~GB memory. It covers a $100\times100~\mathrm{km}^2$ disaster area and includes one satellite backhaul link, three quasi-stationary HAPs at approximately 20~km altitude, twenty sensing UAVs, one surviving emergency gateway, 500 emergency users, and 200 IoT sensors. Each HAP provides 20~MHz sub-6~GHz emergency access, high-capacity Ka-band feeder connectivity, onboard storage, edge computing, AI inference, and solar-powered energy management. UAVs upload visual and thermal observations to HAPs.  Then, HAPs aggregate sensing data, cache emergency maps and AI models, execute local inference for fire-front detection and victim localization, coordinate UAV trajectories, and relay traffic to satellites or surviving gateways. The objective is to maximize the fraction of time during which communication rate, closed-loop delay, sensing quality, data freshness, task-completion ratio, and protected energy reserve simultaneously satisfy the conjunctive availability metric in Eq.~\eqref{eq:s2c2i_availability}. Four representative schemes are compared:

\begin{itemize}
    \item \textbf{Satellite-terrestrial baseline}: Users rely on surviving terrestrial BSs and satellite backhaul. No HAP is deployed.
    \item \textbf{Communication-only HAP}: HAPs provide access and relaying, but sensing data, inference tasks, and content retrieval are mainly handled by the remote cloud.
    \item \textbf{Function-separated HAP}: HAPs are equipped with communication, sensing, storage, and computing modules, but each function is optimized independently.
    \item \textbf{S$^2$C$^2$I-integrated HAP}: HAPs jointly orchestrate sensing rate, cache placement, task offloading, radio-resource allocation, model selection, and energy reservation.
\end{itemize}

Fig.~\ref{fig:case} compares the threshold-satisfaction ratios of the four schemes. The satellite-terrestrial baseline satisfies the rate, latency, sensing-quality, data-freshness, task-completion, and protected-energy requirements for only 45.8\%, 30.6\%, 57.4\%, 41.8\%, 54.9\%, and 60.7\% of the mission horizon, respectively. Deploying communication-only HAPs increases the rate-satisfaction ratio to 90.8\%, which is the highest among all schemes. Nevertheless, its latency and freshness satisfaction ratios remain only 48.5\% and 54.1\%, because raw observations, inference tasks, and uncached contents still traverse the congested feeder link.
The function-separated HAP scheme improves the latency, sensing, freshness, and task-completion performances to 76.4\%, 86.1\%, 81.3\%, and 85.4\%, respectively. However, independent resource allocation causes simultaneous sensing,
computing, and caching modules to compete for onboard power,
reducing its energy-reserve satisfaction ratio to 69.2\%.
In contrast, the S$^2$C$^2$I-integrated HAP achieves
88.7\%, 94.3\%, 92.6\%, 91.8\%, 94.1\%, and 90.9\% across
these dimensions. Compared with the function-separated
scheme, it improves latency, sensing, freshness, task completion, and protected-energy satisfaction by 17.9, 6.5, 10.5, 8.7, and 21.7 percentage points (pp), respectively.

Although joint orchestration slightly reduces the rate-satisfaction ratio by 2.1 percentage points relative to the communication-only HAP, it decreases feeder-link traffic from 1.17~TB to 0.49~TB. It also increases the conjunctive service availability $A_{\mathrm{S^2C^2I}}$ from 34.7\% and 58.6\% under the communication-only and function-separated schemes,
respectively, to 82.4\%. These results demonstrate that the
principal benefit of S$^2$C$^2$I integration is not the isolated maximization of a single function, but the sustained simultaneous satisfaction of communication, sensing, computing, storage, and flight-energy requirements.

\section{Future Research Directions}\label{Sec-VI} 
Although existing studies have demonstrated the feasibility of individual sensing, storage, communication, computing, and intelligence functions on HAPs, a practical S$^2$C$^2$I-integrated HAP-SAGIN must further address autonomous orchestration, cross-domain security, semantic interoperability, and resource-constrained implementation.  Accordingly, this section identifies four promising research directions.
\subsection{Agentic AI and LLM-Native Hierarchical Orchestration}

Recent studies have demonstrated that Large Language Models (LLMs) can be adapted to network prediction, optimization, intent interpretation, and orchestration tasks \cite{liu2025secure}, \cite{luo2024bc4llm}. The evolution from large network models toward agentic AI further introduces planning, memory, tool utilization, reflection, and inter-agent collaboration \cite{luo2026agentic}, \cite{zheng2026agentvne}. These capabilities are well aligned with S$^2$C$^2$I-integrated HAPs, where orchestration decisions must jointly account for sensing urgency, cached-data freshness, radio conditions, computing queues, model accuracy, weather, and protected flight-energy reserves \cite{han2025agent}.

A promising architecture is a hierarchical agent fabric spanning the cloud, HAP, UAV, and ground layers \cite{ibrahim2026survey}. A global cloud agent can interpret long-term mission objectives, retrieve regulatory and network knowledge, and decompose high-level intents into regional tasks. HAP-side agents can maintain regional semantic memories and coordinate sensing rates, cache placement, task offloading, model selection, routing, and energy reservation. Lightweight UAV and ground agents can execute low-latency control through compact models, optimization solvers, or reinforcement-learning policies. Instead of directly generating flight or radio-control actions, LLM agents should invoke verified tools, such as constrained optimizers, network simulators, channel predictors, and safety monitors. This hybrid architecture separates semantic reasoning from deterministic execution and prevents an unconstrained generative model from directly controlling safety-critical functions.

Several fundamental problems remain open. First, agent observations are distributed across heterogeneous and intermittently connected SAGIN nodes, making long-term memory synchronization and state consistency difficult. Second, hallucination, stale retrieved knowledge, erroneous tool selection, and conflicting decisions among HAP agents may lead to unsafe resource allocations \cite{luo2026trustworthy}, \cite{luo2025wbft}. Third, onboard deployment requires model compression, speculative inference, model splitting, semantic caching, and adaptive selection between large cloud models and compact HAP-side models. Finally, agentic orchestration should be evaluated using mission-level metrics rather than language-generation accuracy alone. Relevant metrics include tool-call correctness, constraint-violation probability, planning latency, token and energy consumption, recovery from incorrect actions, and the resulting conjunctive S$^2$C$^2$I availability. Future research should therefore develop bounded agentic autonomy \cite{luo2025wireless}, where every generated plan is grounded in real-time telemetry and verified against communication, computation, storage, energy, and aviation-safety constraints before execution.

\subsection{Security, Privacy, and Trustworthy S$^2$C$^2$I Autonomy}

S$^2$C$^2$I integration significantly enlarges the attack surface of HAP-SAGINs. Sensing functions are vulnerable to spoofing and adversarial observations; storage modules may suffer from cache poisoning and unauthorized model replacement; communication links face jamming, eavesdropping, and impersonation; computing services may be compromised through malicious offloading requests; and intelligent orchestration further introduces model poisoning, prompt injection, unsafe tool invocation, and videsadversarial inter-agent messages \cite{hu2025generative}, \cite{cai2024privacy}. Blockchain provides a complementary decentralized trust substrate for identity authentication, data integrity, access control, and provenance across heterogeneous SAGIN entities \cite{luo2024symbiotic}, \cite{luo2025convergence}. However, most existing solutions still protect individual functions rather than the complete sensing-storage-communication-computing-intelligence decision chain.

A promising architecture is to combine zero-trust access control with a lightweight permissioned blockchain spanning satellites, HAPs, UAVs, and terrestrial centers \cite{jia2026blockchain}. Instead of storing raw sensing data or LLM on-chain, the ledger can retain cryptographic commitments, timestamps, identities, model versions, cache records, offloading decisions, and safety-critical logs. Smart contracts can automate cross-domain authorization, resource transactions, model update approval, and revocation for compromised entities \cite{bikos2025sat}. Since conventional Proof-of-Work (PoW) is infeasible for energy-limited HAPs, future designs should investigate committee-based consensus \cite{liu2025secure1}, hierarchical or sharded ledgers \cite{luo2023esia}, \cite{chen2025drdst}, and asynchronous finality \cite{liu2025asynchronous} to tolerate long delays and intermittently connected SAGIN links. 

However, there are also some problems. First, blockchain can verify provenance and immutability but cannot determine whether authenticated sensing data, AI outputs, or agent decisions are semantically correct \cite{luo2026agentic1}. Second, compromised validators, malicious smart contracts, stale ledger states, metadata leakage, and consensus overhead may introduce new security and resource costs \cite{issa2023blockchain}. Finally, the system should preserve a trusted degraded mode when AI or ledger services become unavailable, retaining command-and-control, emergency access, navigation, and safe-return functions. Security evaluation should therefore include attack-detection accuracy, consensus latency, ledger traffic, recovery time, and the fraction of mission time during which both S$^2$C$^2$I service and security constraints are simultaneously satisfied.

\subsection{Goal-Oriented Semantic S$^2$C$^2$I and Digital Twins}

Most current HAP-SAGIN designs remain bit-oriented: sensing generates raw data, storage retains files, communication transports bits, and computing processes predefined workloads. However, mission-critical applications are ultimately concerned with the usefulness and timeliness of information rather than perfect recovery of every transmitted bit. Semantic and task-oriented communication has therefore emerged as a promising 6G paradigm \cite{sun2024s}, \cite{zhang2026toward}. For S$^2$C$^2$I-integrated HAPs, semantic processing can extend beyond communication and provide a common abstraction across all five functions.

A HAP could determine what should be sensed, which observations should be retained, which semantic features should be transmitted, and where inference should be executed. For instance, during disaster response, raw UAV video may be locally converted into victim locations, confidence values, and hazard boundaries. Only these task-relevant representations need to be cached or transmitted unless uncertainty exceeds a threshold. This approach can simultaneously reduce sensing redundancy, storage occupancy, feeder-link traffic, and computing load. Nevertheless, semantic representations must remain interoperable among heterogeneous satellites, HAPs, UAVs, terrestrial networks, applications, and AI models. Open problems include semantic distortion measurement, multi-task representation reuse, uncertainty calibration, model mismatch, semantic freshness, and graceful degradation under previously unseen events \cite{zhang2026towards}.

Meanwhile, digital twins can provide the environment for developing and validating such goal-oriented operations. Existing research has introduced digital-twin channels, network digital twins, and resource allocation based on virtual replicas of NTNs \cite{wang2025digital}, \cite{al2024digital}. A HAP-oriented twin should jointly reproduce platform motion, atmospheric conditions, RF/optical channels, propulsion and solar energy, thermal states, computing queues, storage contents, AI-model behavior, and service demand. However, continuously synchronizing a full-fidelity twin may itself consume excessive sensing, communication, storage, and computing resources. A more scalable direction is the goal-oriented semantic twin \cite{qiu2025twinning}, which retains only the state variables and causal relationships required by the current mission. Research is needed on adaptive twin fidelity, physics-informed learning, uncertainty propagation, causal what-if analysis, online twin calibration, and transfer from simulation to stratospheric flight. Ultimately, the twin should not merely visualize the HAP network; it should predict the effect of candidate S$^2$C$^2$I actions and verify them before real-world execution.

\subsection{Sustainable, Certifiable, and Open HAP-Native Systems}

The resource assumptions commonly adopted in terrestrial edge computing cannot be directly transferred to HAPs. Onboard intelligence competes with sensing, baseband processing, communication, storage, propulsion, and thermal control for a limited energy and payload envelope. Green edge AI research has emphasized the importance of jointly considering model accuracy, inference latency, computation, communication, and energy \cite{mao2024green}. More recent studies investigate token-responsive energy management and the integration of TinyML and large models across device-edge-cloud hierarchies \cite{yu2025tree,vu2025tinylarge, luo2024communication}. These ideas should be extended to a HAP-native computing architecture. Compact models handle frequent real-time decisions, while larger models are selectively activated, partitioned across layers, or accessed through intermittent satellite and terrestrial links.

Future work should jointly design heterogeneous onboard accelerators, model compression, quantization, dynamic voltage and frequency scaling, memory and cache hierarchies, thermal-aware scheduling, and function migration. The relevant objective is not merely bits/J or inference/J, but total mission energy per completed S$^2$C$^2$I task, including propulsion, sensing, storage refresh, model transfer, radio transmission, and fault recovery \cite{luo2026agentichaps}. Resource managers must also reserve sufficient energy and computing capacity for command-and-control and safe flight rather than allocating the full payload budget to commercial services.

Openness and certifiability are equally important. AI-native O-RAN has been proposed as a means of introducing disaggregation, virtualization, and intelligent orchestration into NTNs \cite{deng2026oran}. However, HAP systems still lack common descriptors for sensing payloads, accelerators, storage state, energy reserves, environmental conditions, and mission constraints. Future platforms should support lightweight containers, standardized telemetry, portable xApps/rApps, secure update and rollback, fault containment, and multi-vendor interoperability. At the same time, adaptive AI components must generate evidence suitable for aviation and telecommunication certification, including bounded execution times, verified resource limits, and reproducible failure tests. Open HAP digital twins, environmental traces, and anonymized flight datasets are therefore essential. This would enable the community to distinguish algorithmic improvement from practically deployable S$^2$C$^2$I capability.

\section{Conclusion}\label{Sec-VII} 

This survey has examined the evolution of HAPs from communication relays into multifunctional infrastructures for future SAGINs. We introduced a unified S$^2$C$^2$I perspective in which sensing, storage, communication, computing, and intelligence are jointly designed under stringent constraints. We further reviewed the roles of HAPs, the cloud-edge-HAP space continuum, and key enabling technologies, including heterogeneous connectivity, ISAC, onboard computing, caching, and AI-based orchestration. The proposed evaluation framework and emergency-response case study demonstrate that cross-functional coordination can improve service availability, reduce feeder-link traffic, and preserve flight-energy reserves.

Despite recent progress, practical deployment still requires trustworthy autonomy, cross-layer security, interoperable interfaces, mission-aware digital twins, and energy-conscious onboard intelligence. Reproducible benchmarks and realistic testbeds are also needed to jointly evaluate communication, sensing, computing, storage, intelligence, and flight safety. Ultimately, S$^2$C$^2$I-integrated HAPs can serve as adaptive regional intelligence layers that transform heterogeneous observations and distributed resources into timely services.
\bibliographystyle{IEEEtran}
\bibliography{IEEEabrv,mylib}








\vfill
\end{document}